\documentclass[aps,prd,nofootinbib,preprint,tightenlines]{revtex4-1}
\DeclareFontFamily{U}{rsfs}{}         
\DeclareFontShape{U}{rsfs}{m}{n}{<5> rsfs5 <6><7> rsfs7          %
  <8><9><10><10.95><12><14.4><17.28><20.74><24.88> rsfs10}{}     %
\DeclareMathAlphabet{\mathfs}{U}{rsfs}{m}{n}                     %
\usepackage{amsmath,bbm}
\usepackage{amssymb}

\def\be{\nopagebreak[3]\begin{equation}}
\def\ee{\end{equation}}
\def\ba{\nopagebreak[3]\begin{eqnarray}}
\def\ea{\end{eqnarray}}

\def\l{\langle}
\def\r{\rangle}

\newcommand{\teta}{\rlap{\lower2ex\hbox{$\,\tilde{}$}}\eta{}}

\newcounter{mnotecount}[section]

\begin{document}
\preprint{\vbox{\baselineskip=12pt \rightline{ICN-UNAM-05/01}
\rightline{gr-qc/yymmnnn} }}
\title{ Shortcomings in  the Understanding of Why Cosmological  Perturbations Look Classical}

\author{Daniel Sudarsky \thanks{sudarsky@nucleares.unam.mx}}
\address{  Instituto de Ciencias Nucleares\\
  Universidad Nacional Aut\'onoma de M\'exico\\
  A. Postal 70-543, M\'exico D.F. 04510, M\'exico\\
}

\begin{abstract}
 There is a persistent state of confusion regarding the account of the quantum origin of the seeds  of cosmological 
 structure  during inflation.  The issue  is  the  transition  from  the quantum uncertainties in the homogeneous and isotropic initial state, into the  late time  ``classical" anisotropies 
 and inhomogeneities. There seems  to be a  widespread  belief that decoherence  
addresses  the issue  in a  satisfactory way.  This  view is taken, often implicitly,  by  most  researchers  working in the field. This  can  be seen most  clearly in  those   accounts  intent on 
 facing the issue  directly.  
  For instance, a recent article 
  \cite{Polarski-Kiefer} argues  just  that, and presents  a detailed  explanation of the justifications. The  explicit  nature of that account  will allow us  to  discuss the issue   in  detail.
  There  are, of course,  various  other  works  that often indirectly   address  the issue with similar approaches \cite{Cosmologists}.
This type of  arguments, do not only implicitly  assume  that decoherence  offers  a satisfactory  solution to the measurement
 problem in quantum mechanics, but also that,  in particular,  such approach is  applicable  to these quantum aspects of  cosmology. We will review here, why do we, together with various  other researchers in the field, consider 
that this  is not the case in general. Moreover as  
has been  previously discussed \cite{Us, othersus},  we will  argue that the cosmological  situation 
is one  where the measurement problem  of quantum mechanics appears  in a particular exacerbated form,  
and that, it is this,  even  sharper  conundrum,  the one that should be  addressed when dealing with the inflationary account of 
 the origin of the seeds of  cosmic structure  in the early universe. 
 We  briefly discus   the ideas behind 
 what we feel might be a promising approach.
\end{abstract}

 \maketitle

\section{Introduction}

  The coming of age of cosmological scientific inquiry  with the recent outpouring of  high precision empirical data, has been accompanied with   the  establishment of the inflationary paradigm as part of the standard model of cosmology. In fact,  one  of the major  successes of inflationary cosmology is considered to be  its  ability 
to ``account for " the spectrum  of the temperature anisotropies in the cosmic microwave background, which is
understood as the earliest observational data about  the primordial density fluctuations that seed the
growth of structure in our Universe. 

The impact of the data obtained by the various missions: COBE, Boomerang, WMAP, and, hopefully in the near future, 
Planck, on our understanding of the universe, reaches up to  the very sources of the processes that made us possible: Galaxies, Stars, Planets, life and human beings  with all their scientific advances should be regarded as results of the evolution of those early inhomogeneities we are now 
studying. 

However,
when considering this account in more detail,  one immediately notes that there 
is something exceedingly strange
 about it, namely, that out of an initial situation, which is taken to be  perfectly homogeneous and  isotropic   (H.\&I.),  
and based on a dynamics that supposedly  preserves
 those symmetries, one ends with a non-homogeneous and non isotropic situation.  Most of our colleagues, who have been working in this 
field for a long time, would reassure us,  that there is no problem at all by
invoking a variety of arguments.
 It is noteworthy  that these arguments  would tend to differ, in general, from one inflationary cosmologist
 to another \cite{Cosmologists}.  Other cosmologists do acknowledge that there seems to be something unclear at this point 
 (\cite{Padmanabhan} for instance,  indicates that one  must  work  with {\it certain classical objects  mimicking the  quantum fluctuations},  
 and that  this is not easy to do  and  to justify). Actually,  some recent books  on the  subject  acknowledge  that there is a  problem.  For instance, in \cite{Weinberg}, 
  we find {\it ``...  the field  configurations must become   locked  into  one  of an ensemble of classical configurations  with ensemble  averages  given by  quantum  expectation values...
    It is not apparent  just how  this  happens...."}, while  \cite {Mukhanov} clearly acknowledges that  the problem is not resolved  
 simply by  invoking decoherence:  {\it `` ..  However decoherence is not enough to explain the breakdown  of translational invariance.."}. Nevertheless most  of  the cosmology community continues to hold the belief that the issues  have been successfully resolved within that approach.
 
 In a recent  series of  papers \cite{Us, othersus},
a critical analysis of such proposals has been carried out indicating that all the existing  justifications fail to be fully satisfactory.

    The  aim  of this  manuscript is  to  present, in general terms,  the problems surrounding the  accounts of the quantum origin of cosmic  structure, placing the   issue  within the general  conceptual framework concerning the  interpretation of quantum theory in general,
    and, to illustrate the  various traps,  sources of confusion and  inconsistencies  one  might encounter in this enterprise.  We  will focus  on the approach   most  widely accepted  in the   community  working in the field, and  explore the shortcomings through  the discussion of  a specific  example.  Recognizing the extent of these  shortcomings, is the first step in  the uncovering of what might  be a thread that connects  the problem posed by the quantum origin of the  cosmological structure  with other conceptual  and technical problems,   facing  our current theoretical framework,
    such as the   general measurement problem in quantum theory, or the problem of time, which  arises in  many  proposals for   putting together quantum theory and gravitation.
  The article is  organized as follows: In Section II  we review  the   general setting,  concentrating on the  relation between the classical and  quantum descriptions of  a  problem,  and  the difficulties in the  transition between those two in the cosmological context in particular. Section III,  will treat the applicability of  quantum theory to cosmology, within the framework provided by the general  views  one  can take  about   quantum mechanics. Section IV,  will focus on the decoherence approach and  refute  the  widespread notion that it,  by itself, fully  addresses the problem  at hand. In Section V,  we will argue that  the measurement problem becomes  even more vexing in the cosmological context.  Finally,  in  Section VI,  we identify the  specific  manner in  which  problems  appear in specific  accounts  by focussing  on a particular  work \cite{Polarski-Kiefer}  which  we consider as  characterizing  the  predominant view  regarding the issue,  and   show  the extensive set of  problems  that are overlooked  when  taking  such posture. Section  VII   will briefly discuss other approaches that have been  considered  for dealing   with the problem. Section VIII  presents  an overview  of an approach,  implemented  explicitly in \cite {Us, othersus, Adolfo}, that    we have  been  advocating   and that  seems  capable of successfully addressing  the issue, at the price of adding new physics, in the form of a  dynamical collapse of the wave function, inspired  by  Penrose's  ideas.  There the issue of how  could such dynamical   reduction   be incorporated in the  general   current paradigm  of physics is  addressed  in a rather  schematic  way, hoping that it may  serve as  the basis  for the development of a more rigorous   treatment.      We   end  with a  brief conclusion. 
  
 \section{The Problem}
 
For most of its existence, cosmology has been discussed in a classical language, as  it is in fact done in 
many other disciplines  closely tied to  physics, such as  hydrodynamics,  bridge building,  and  the study
  of trajectories of space probes, while everybody knows that our world is quantum mechanical\footnote{Throughout this  manuscript , we  will take the conservative view that  Quantum Theory has  universal validity, unless something  else is stated  specifically  within certain specific discussions  of alternative  views.}.
   The justification is, of course,  that  we believe that the classical description is nothing but an approximation to the quantum description,   and so, when we consider, say, the classical  description of the  trajectory of a satellite in space, we view it as   indicating that the wave function of its constituting atoms  (or even that of  its  more  elementary constituents) is a sharply  peaked wave packet where the uncertainties in the  position and velocities are negligible compared with the  precision of the description we are making. The classical  description  does not enter into a fundamental contradiction with the characteristics of  our satellite trajectory.  It would be very unsettling if we were forced to consider the classical  elliptical trajectory  around Earth  while at the same time  we  were forced to admit that, at  the more fundamental level, the satellite was described by a   spherically symmetric wave function.  We know this is not the case,  because it is clear that  the precise  quantum description  of the situation would call for a suitable superposition of   energy and momentum eigenfunctions leading to a  wave packet corresponding to a sharply localized object. Needless is to say that the  precise way to accomplish this  is filled with issues that, at this time, are  technically insurmountable, but this is not  the point here, the principle is clear. Moreover,  we should recognize that in the case of the  satellite, one is  dealing with an open system, and its interaction with the clearly identifiable environment is  likely to play an important  role in   making compatible  the quantum and classical descriptions.

In the cosmological setting, however,  when we  want to connect  our  classical descriptions  of the cosmological late times with a  quantum description of early cosmological  times, we cannot  escape from the corresponding requirements.  We know that our universe is quantum  mechanical and, thus, the classical descriptions must be  regarded as nothing but short-hand  and imprecise characterizations of complicated quantum mechanical states involving peaked wave packets  and  complex correlations. The universe that we inhabit today is certainly  very well described  at the classical level by an in-homogeneous and anisotropic classical state, and  such  description must, in accordance with the previous paragraph,  be nothing but a concise
and imperfect description of an equally in-homogeneous  and anisotropic quantum state, where the wave functions are  
peaked  at those  values of the  variables  corresponding to   those  indicated  by the  classical description (or  values very close to those).  This would, in principle, involve no essential differences from the  case of the classical and quantum description of our satellite, except for the lack of a clearly identifiable  environment, if we take the universe to include,  by definition, all the degrees of freedom of our theory.   But there is nothing that indicates that, even without the identification of an  environment, we should  not be able to make, in principle,  such  quantum  semi-classical description  through the use of the sharply peaked wave functions and taking into account all the  interactions in the analysis of its  dynamics.  The situation changes dramatically,  however, if we want to  seriously consider a theory, in  which the early  quantum state of the universe was particularly simple in a very special and  precise way.  
 This is the case in the inflationary paradigm, and  in particular,  as it refers to the predictions about the spectrum of  perturbations that  we believe constitute  the seeds of cosmic structure.

 Inflation was initially devised as a mechanism to deal  with the, so called,  naturalness problems of standard Big Bang Cosmology:  Namely, the Horizon problem,  The Flatness problem  and the  Primordial relics problem \cite{Problems Big Bang}.  The essential  idea is that  if the Universe undergoes an early era of  accelerated  (almost exponential) expansion (lasting  at least some 80 e-folds or so), it would come out of this period as an essentially
 flat  and homogeneous space-time  with an extreme  dilution of all relics  and, indeed, of  all particle species.  The states of
  all fields  would,  thus,  be  extremely well described by  suitable  vacua.
 The deviations from this state will be exponentially small.  
 What is needed is something that behaves early on as a cosmological constant but that is later
``turned off" as a result of its own dynamics,  returning the universe to the standard Big Bang cosmological evolutionary path. This is  generically thought to be  the result of a scalar field  with a potential  of certain specific characteristics  called the ``inflaton field".
The remarkable fact is that this scheme also
 results in a spectrum of primordial  quantum uncertainties of the inflaton field that matches the form of the famous Harrison-Z{}'eldovich spectrum of primordial perturbations and which has been observed in the  various analysis of  the  extraordinary data on the CMB sky collected in the various recent experiments.

This is the basis of the claim that inflation ``accounts for the seeds of the cosmic structure".  They ``emerge from the quantum vacuum", continue to evolve  after inflation has ended, and  after  leaving   their mark on the CMB, result in the emergence of  the structure  of our universe.  That  structure which  at late  times is  characterized by 
the conditions permitting our own existence.

 The predictive power is, thus, remarkable, but it would be completely lost if we needed to rely on the imperfections of the  state that resulted from inflation. 
 That is, if the whole picture  depended,  in a critical way, on the minute deviations from the very simple state that we indicated  is thought to  
 describe the universe as it emerges from  the inflationary regime,  we would not be able to make any predictions at all.  
 It could well be that the structure  of  our universe today  (and the markings on the CMB)  depended in that way 
   on those  minute inhomogeneities,  but in that event, the matching of the spectra of the quantum uncertainties and that of the CMB spectra would be  nothing but a miraculous coincidence. We do not believe  in such coincidences;   however,  that  does not mean that we should avoid taking a critical look at the  ``successful account".  In particular, we must  carefully inquire into the identification of these quantum uncertainties and the
 perturbation spectrum  for the seeds of structure that characterize our universe.    In so doing, one must  
 be careful in  taking  the simple state  that inflation provides as if that was the complete  description of the starting point of the whole  account of the origin of these  fluctuations, because  we have no justification for  claiming anything about the remaining imperfections, except  that they are exponentially small.
 
Therefore, in turning back to the main subject,
  the issue  we want to discuss is: does the standard inflationary scenario,  which starts off by arguing that the universe entered, as the result of inflation, into a simple state characterized as the vacuum of all fields and the flat FRW space-time,   when put together with  decoherence arguments,  really account for the  transition from that  H.\&I. early state of the universe,  to the  anisotropic and inhomogeneous universe we inhabit\footnote{This point is sometimes characterized as the ``transition from the quantum regime to the classical regime", but we find this a bit misleading: most people would agree that  there are  no classical  or quantum regimes. The fundamental description ought to be always a  quantum description. However,  there exist regimes in which certain quantities can be  described  to a  sufficient accuracy by their classical counterparts represented  by  the  corresponding expectation values.   All this depends, of course, on the physical state, the underlying dynamics, the quantity of interest, and the context in which we might want to use it. 
  }?

  This article will be devoted, to a large extent, to deal with the conceptual  issues above, and will  
  not include the developments that are possible   when taking  a stricter ontological view of
  the essence of quantum physics,   and consequently adding new  elements to deal with the shortcomings  we encounter in the present context, and   would refer the interested readers to other articles that dwell on those topics\cite{Us, othersus}.
  
We now  proceed to  consider the views  one can take regarding    quantum theory, and  how  they impact  its application to  
problems  in cosmology .
    
\section{ Cosmology and Views  on Quantum Physics }

In analyzing the present topic,  we have  to consider the  views  that one might take regarding  cosmology and quantum physics, first to enhance the sharpness of the discussion,   in order to avoid   possible  misunderstandings,  and to  clarify to the reader the reasons for our posture regarding the matter at hand.  
    
       The issue  we are facing is  related to the so called  ``measurement problem in quantum mechanics", a subject that has puzzled physicists and philosophers of physics  from the time of the inception of the theory  \cite{MPQM}.   These issues continue to  attract the attention  of   several  thinkers in our field,  and  we  are, for the most part, merely recounting  the status of the general  problem, touching,  when appropriate,   on the  particular  instance   that concerns us here:  the cosmological setting,  a subject that has received  much less attention from the physics community.    There are, of course, notable exceptions to the assessment above, represented by  thinkers like R. Penrose\cite{Penrose1},  J. Hartle \cite{Hartle1},  and others.

     Any reasonably complete discussion of the  interpretation of Quantum Mechanics  is well beyond the scope of this manuscript,  and needless is  to say that much more  exhaustive   discussions on the subject  exist in the literature \cite{Nuevos  QM}, but  we  must  include  a very  brief account of what  seems to be  the most common  postures, in order to put  our discussion in the appropriate context. 
Let us,  thus,  briefly  consider  some  of the  views that can  be taken on the  subject of Quantum Theory  and the ``measurement problem" that are relevant to our situation:

\begin{itemize}
  
  \item {\bf a) } {\it Quantum physics  as  a complicated theory of  statistical physics.}
     It is a  position that holds that quantum mechanics acquires meaning only as it is applied to  an ensemble of  identically prepared  systems.  In this view, one must accept that a single atom, in isolation, is not described by quantum mechanics.
 Let us not get confused by the correct, but simply distracting, argument that atoms in isolation do not exist.  The point is whether,    to the extent to which we do  neglect its interactions with distant atoms,  and specially  with the electromagnetic field which, even in its vacuum state is known to interact  with  the atom, quantum mechanics is applicable to the description of a single atom. Again, what can we mean by that, if we know that, in order to be able to say anything about the atom, we must   make it interact with a  measuring device?  Well, the question is simply whether applying the formalism of quantum mechanics  to treat the isolated atom can be   expected to yield correct results  as it pertains the subsequent measurements.?  One might  think that this is a nonsensical question, as these results are always statistical in nature. The point is that this  statement is not really accurate: for instance, if the atom (say, of  hydrogen) was known to  having been prepared  in its ground state, the probability of measuring any energy, other than the one in the ground state, is zero. 
 In fact, for any observable commuting with the hamiltonian  the predictions are not statistical 
        at all, but 100\% deterministic and precise! If so, there must be something to the description of that single atom by its usual  quantum mechanical state. 
  It is  thus   a clearly  false  notion, the  idea that quantum mechanics can not be applied to  a single 
  system \cite{Single System}. What is true, of course,  is that, in applying the theory to a single system, the  predictions we  can make with certainty are very  limited,  with the extent of such limitations being determined, both by the nature of the system's   dynamics, and by the way the system was initially prepared. Moreover, in relation with the issue that concerns us in this article, taking  a posture
 like this about quantum physics,  would be admitting  from the beginning that we  would have no justification in   employing such theory in addressing questions concerning the  unique
 universe to which we have access, even if we were to accept that somehow there exists an ensemble of universes  to which we have no access. Note, moreover, that  we should beware from confusing statistics of universes and statistics within one universe.  Furthermore, if a quantum state serves only to represent an ensemble,  how is  each element of the ensemble to be described?  Perhaps, it can not  be described at all ? What do we do in that case with our universe?
      
\item      {\bf b) }   {\it Quantum physics as  a theory of human knowledge}. Within 
     this  view of quantum theory,  the state of a quantum system does not reflect something about the system, but just what we know about the system\footnote{One can find  statements in this  sense  in well known books, for instance ``Quantum Theory is not a theory about reality, it is a prescription for making the best possible predictions about
 the future, if we have certain information about the past" \cite{Asher Peres}.  See also\cite{Hartle-objective}}.  Such view, naturally rises the question: what is there to be known about the system if not something that pertains to the system?  The answer comes in the form  of: correlations between the system and  the measuring devices, but then,  what is the meaning of these correlations? The usual meaning of  the word ``correlation" implies some sort  of  coincidence of certain conditions pertaining  to one object with some other conditions pertaining to the second object.  
However, if a quantum state describes such correlation,  there must be  some meaning 
 to the conditions pertaining to  each  one  of the objects. 
         Are not these, then,  those very same aspects that are described by the quantum mechanical state for the object? If we answer in the negative,  it must mean that there  are further descriptions of the object that can not be casted in the quantum mechanical state vector. On the other hand,  if we answer  in the positive we are again taking  a view whereby the state vector says something about the object in itself. Perhaps,   we are just going in circles. If one  were inclined to  read these  considerations as philosophical  nonsense, one  should not forget that if we  follow the above described  view, we  would have  abandoned the possibility to consider  questions about the evolution of the universe in the absence of sapient beings, and much less to consider the emergence, in that universe,  of the conditions that are necessary for the eventual evolution of humans, while using Quantum Theory. We could take it even further and ask ourselves: what would be, for instance,  the justification for considering  states in any model of  quantum gravity, if we took such view of quantum physics?

\item      {\bf c) }  {\it Quantum physics  as  a non-completable  description of the world.}  Here,  we   refer to any posture that effectively, if not explicitly,  states: ``The theory is incomplete, and no complete theory containing it exists or will ever do". Such  view  will  be  considered as  being advocated  by  any posture indicating, either  openly or implicitly,  to use quantum mechanics ``as we all know how",   and supported by the observation that no violation of quantum mechanics has ever been observed\footnote{In practice,  this view is essentially indistinguishable from the so called FAPP ({\it For all Practical purposes}) approach to the matter \cite{FAAP}}.  While this  is, with not  doubt, a  literally correct statement, we must remind our colleagues that by this,  
one  would refer, of course,  to the rules as found in any quantum mechanics  text book, that  essentially rely on the Copenhagen interpretation, which, as we all know,  raises  severe  interpretational issues  that  {\bf become insurmountable  once we leave the laboratory and consider applying quantum theory to something  like the universe itself}.  According to this view, we should content ourselves using its tools, and making, in the situation at hand,  ``non-rigorous predictions".
  We must acknowledge, however, that in  situations where one  can not point to the classical-quantum dividing line, where we can not identify the system and the apparatus, nor the observables that are to be measured, the entity carrying out those  measurements and the time  at which the measurements are to be thought as taking place, we have no clearly defined scheme specifying  how to make the desired predictions. That is, in dealing with the  questions pertaining to the early universe in terms of quantum theory, we  have no clear and specific rules for making predictions.  However, according to such  practical posture, we should be content  with the fact that the predictions have, in fact, been made, and that they do seem to agree with observations. The issue is, of course, that in the absence of a well defined set of rules,--- rules that are explicit to the point where a computer could,  in principle, arrive to the predictions using only the explicit algorithm and the explicitly stated inputs--, we have no way to ascertain whether or not,  such ``predictions" do  or do not,   follow  from the theory.  We can not be sure whether or not, some unjustified  choices, manipulations  and arguments have   been used as part of  the process by which the predictions have been obtained. {\bf Correct quantitative} forecasts are not  enough. These must first be  {\bf actual predictions}.  This should be
  quite clear, particularly, when  thinking about  cases where the argumentative connections used in arriving to the ``predictions"  are so loose that never-ending debates can  arise  which can not be translated into  arguments about   precise  mathematical statements.  Specially suspicious are,  of course, those ``predictions" which  are, in fact, retrodictions, and on this point we should be aware that long before inflation  was invented,  Harrison and Z{}'eldovich \cite{HZ} had already concluded  what should be the form of the primordial spectrum, based  on rather broad observations about the nature of the large scale structure of our universe. 
      
    \item     {\bf d)} {\it Quantum physics as  part of a  more complete  description of the world.}  Here we are not referring to an  extension of the theory  into some sort of hidden variable type,  as the problem we want to face here  about  the theory  is not its indeterminism, but the s``measurement problem" (see \cite{Measurement}  for  more extensive discussions on the problem). 
 Completing  the theory would require something  that removes the need for a  external measurement apparatus, an external observer, etc.  
      Here we  must include D. Bohm«s  `` Pilot Wave Theory"\cite{Bohm}, and   in particular we should mention a specific  proposal to  apply  such ideas to the cosmological problem at  hand\cite{Valentini}. We  will  briefly touch on  the  difficulties  we see in it, in section  VII.   Then there are 
 other  proposals invoking something like the  dynamical reduction models proposed   by Ghirardi, Rimini \& Weber \cite{GRW}, and 
  the ideas of   R. Penrose about the role of gravitation in modifying quantum mechanics in the
  merging of the two aspects of physical reality \cite{Penrose1} (See also \cite{Anandan}).  In the  particular context of inflationary cosmology, our own work \cite{Us, othersus, Adolfo} offers an example of an analysis  in which the issues are faced  directly, and which leads to the possibility of confrontation with observations.  In this  sense,  this  is the position we advocate, inspired  in part by the arguments made in\cite{Penrose1, GRW}, and  by the problem at hand.

\item         {\bf  e) } {\it Quantum physics as  a complete  description of the world.}   The view that quantum mechanics faces no open issues and that, in particular, the measurement problem  has been solved.   Among the holders of these views one can further identify two main currents:  those that  subscribe  to  ideas along the so called  "many world interpretation of quantum  mechanics" and consider this to be a solution to the measurement problem, and  those that hold a view  that the measurement problem in quantum mechanics has been solved by  the consideration of    "decoherence".  Let us first note that the many world interpretation does very little to ameliorate the measurement problem, as there is a mapping between what in that approach would be called  the ``splittings of worlds"\footnote{It often claimed   that  there are no  "splitting of the  world  " in the many world  interpretation, but the  fact of the matter is that  whenever people  make use of it they can not avoid talking  about   things  such as "our branch", "the realms  we perceive" or other  notions  that   implicitly  make  use of  a notion  that  is  essentially just that  a "world  splitting". One can  see this in each specific  application  of the idea,   by focussing on the complete description of what one would take as " the relevant state  describing our  reality"  and following it in time  backwards and forwards.  In the  inflationary situation at hand, this is  easily done,  by focussing on the symmetry of the state  describing the quantum fields.} and what  would be  called  "measurements" in the Copenhagen interpretation. Thus every question that can be made in the latter interpretation  has a corresponding one in the many worlds interpretation.  For the case of the measurement problem the issues would be:  When does a  world splitting occur? Why, and under what circumstances does it occur?   What constitutes a  trigger? 
   Finally   we should mention the  proposal  for   generalization of quantum physics using a scheme  based on   the realms of decoherent  coarse grained  histories  proposed by J. Hartle  \cite{Hartle1},   and  related ones, 
   an approach  that we  will consider  briefly  in section VII.
 \end{itemize}       
           
 Next, as it seems  to be  the  most  widely held  view  on the   way of addressing our problem,   we shall consider  the  decoherence  ``solution" in more  detail.  
 
 \section{Decoherence}         
          
       Decoherence is  a clear and inescapable prediction of quantum mechanics and it has important implications  in  many experimental situations, particularly  in the design of potentially useful quantum computers.  However,  there is  a   widespread belief that it has  implications that  go  well  beyond  and allow s for a complete and satisfactory solution of the measurement problem in quantum mechanics.
       This  is not the case,  but unfortunately the  exhaustive discussion of this issue  alone  would  require  an inordinate  amount of space.  The reader is thus   directed to consult the vast literature on the matter  (including  that  referenced in this  manuscript ) and   here we  will provide only a small  ``taste" for  the issues. For a sharp  discussion   on the  subject  we   refer the reader  to \cite{Adler}.
        We should  start  by quoting the postures that in this regard are held by  several people that have considered the issue at length, in order to  contrast  them  with the often  held  notion that such is the consensus view:

      Take for instance  the conclusion:
    {\it  ``Many physicist nowadays think that decoherence provides a fully satisfying answer to the measurement problem. But this is an illusion."} Arnold Neumaier \cite{Neumaier}.

      Or the warning against misinterpretations:
      
    {\it `` ...note that the formal identification of the reduced density matrix with a mixed state density matrix is easily misinterpreted as implying that the state of the system can be viewed as mixed too..
   ..  the total composite system is still described by a superposition, it follows from the rules of quantum mechanics that no individual  definite state can be attributed to one of (the parts) of the system ..."}, M. Schlosshauer \cite{Schlosshauer}.

      Or the  explicit refutation:
      
    {\it  ``Does decoherence solve the measurement problem? Clearly not.  What decoherence tells us is that certain objects appear classical when observed, But what is an observation? At some stage we still have to apply the usual probability rules of Quantum Theory."}  E. Joos  \cite{Joos}.

We  will, of course, not reproduce the  complete discussion by these authors here   and would turn the interested reader to the corresponding references.

   However, we will see that, when dealing with cosmology, the problem becomes even more vexing and acute.  
 We should note  however, that   there exists   in the field a widespread notion  that  some version of decoherence provides   the paradigm where the  direct application of  the standard forms of quantum mechanics to the problem  at hand finds its justification. Significantly, the diversity of precise approaches   indicates also a   degree of disagreement in the field,  about    the details of such  schemes\cite{Cosmologists}.    
   
 Before engaging on the specific discussion of  the cosmological case,  let us review briefly what decoherence is, and what it can and cannot do.

  \subsection{Decoherence and the   measurement problem}
  
    Decoherence is the process by which a system that is not isolated, but in interaction with an environment (as are all physical systems, except the universe itself) ``looses" or ``transfers" coherence into the degrees of  freedom of such environment. It is a well studied effect that follows  rather than supersedes  the laws of quantum physics.  It is, therefore, clear that, in principle, it can not be thought to offer  explanations  that  go beyond what  can be directly inferred from the application 
    of the principles of quantum physics. Its main achievement is to allow for the study of
     the conditions in which the quantum interferences expected 
    from the idealized  consideration of a system as isolated, become observationally suppressed as the result of the system's  interaction with the environment. 
    
The basic  recipe for an analysis of decoherence in a given situation follows the following steps: 
\begin{itemize}
   
\smallskip 
 \item  {\bf  1)} Divide D.O.F. : system + environment
  ( identify inaccessible or irrelevant D.O.F.).
  
\smallskip
   \item  {\bf 2)} Compute  Reduced Density matrix
    (trace over environment D.O.F.).
    
\smallskip
  \item   {\bf 3)} Perform suitable time average so that the off-diagonal matrix elements vanish.

\smallskip
 \item    {\bf 4)} Regard the diagonal density matrix as describing a statistical ensemble.
 \end{itemize}    
\bigskip    
    
    {\bf The Problems:}  once one has understood  why  certain interferences can not be observed  in practice, it is tempting to conclude that one has understood the ``emergence of classicality", and 
    that, therefore, there is nothing  left of the so called  ``measurement problem" in quantum mechanics. This  turns out to be  a simplistic  and misguided  conclusion, as indicated by the quotations listed  above. There are, actually,  at least two  very serious problems  with considering decoherence  in this light: 
   
   \begin{itemize} 
   \item     {\bf  I) The  basis problem:} it is clear that the diagonal nature of the  reduced density matrix, obtained in the step 3) 
    of the  program above,  will be lost, in general, upon a change of basis for the Hilbert space of the system at hand. 
     This is taken  to mean that the nature of the  system-environment interaction selects a so called pointer basis, 
     which underscores the aspects  that have become classical as  a result of decoherence. The point, of course, is that
      this leaves one with the usual situation whereby, if the selected basis is, say,  the position basis, or one  made  out of narrow wave packets in  position space,  the momentum
       of the system  would remain, correspondingly, highly undetermined, and thus one could not argue that classicality has really emerged.  
       It is  only  that the uncertainties might be  small "for all practical purposes"    something that underscores  
          the need to  further   specify those purposes, and  which  necessarily refers us  back to   certain practicalities of the human condition.

      \item    {\bf II) The definite outcomes problem: } here the problem is  the absence of sufficient justification for the 
   interpretation of the mixed state described by the density matrix as describing   a statistical ensemble,  and in regarding  a single system as being in  a  definite, yet unknown, state among the ones 
   represented in the diagonal elements of the density matrix.  The result  that emerges  from the decoherence  
   calculations rather  indicates that the system must be regarded as coexisting in
    the various alternatives, but with the interferences in  the  appropriate observables, being suppressed.  Selecting among these   alternatives  can be viewed  as deciding  between  
   the ``choice vs. coexistence" interpretations. In order to argue that decoherence really leads to the emergence of classicality, one would have to advocate  the ``choice" interpretation; i.e.,  that somehow  only one of the  decoherent possibilities represents reality, but  well known, and experimentally confirmed aspects of quantum mechanics  such as  the violation of Bell's inequalities, force us to opt for the ``coexistence"   interpretation\footnote{In an Einstein-Podolsky-Rosen-Bohm Gedankenexperiment (EPR) setup, for instance, the experimentalist who is dealing with one of the particles of the  EPR pair,  might invoke decoherence by arguing  that the D.O.F. of the second particle are, at that time, inaccessible to him.  But  taking this as support for the  ``choice"  view  would lead  to the ``conclusion" that the  individual particles have (even if unknown to him) a  well defined spin. This  conclusion   is incorrect, as  it is known to lead to contradictions (See\cite{Mermin}).}\cite{Bell}.
\end{itemize}         

The next example,  from ordinary non-relativistic  quantum mechanics,  serves as a clear analogy of the situation we face:
consider a single particle in a state corresponding to a  minimal  wave  packet centered 
 at $ \vec X = ( D, 0, 0)$ ( the vectors in 3-D space are given in cartesian coordinates  (x, y, z)).
  Let the particle have its spin pointing in the $+y$ direction.  Take this state and  rotate it by and angle $\pi$ about the $z$ axis.  
  Now consider the superposition of the initial and the rotated states. The resulting state is clearly  symmetric under rotations by $\pi$ around $z$.
         Now consider taking the trace  over the spin D.O.F.,  simply because  we  decide  to ignore   the spin variable (no measurement is contemplated here).
The resulting density matrix is diagonal.
 Can we say that  the situation  has become classical? Of course not.   Is the state still invariant under rotations of 
magnitude $\pi$ about the $z$ axis? Obviously, the  answer is  yes.
Can a mathematical manipulation with no physical 
process counterpart (as  ignoring  some  D.O.F.  would  normally not be thought as having an effect on the system) ever change the state of the system? Answering yes would take you to the view discussed in b) of section III.

Therefore, we must conclude that, neither  decoherence, nor the many worlds interpretation do offer satisfactory solutions to the measurement problem.  But, is it perhaps the case that  when both are put together   they   
do offer one?  We  face  that  question  next.

\subsection{On the shortcomings  of decoherence plus many world interpretation as solution to the measurement   problem}

  Consider the traditional  problem  of Shroedinger's cat: The traditional paradigmatic problem that
   brings to the forefront the interpretational problems of Quantum Mechanics. A cat is placed in a box  with an excited  atom  whose decay is arranged to trigger the release of certain  poisonous gas that would kill the cat.  The unitary evolution of Quantum Mechanics then leads, after some appropriate time,   to a  state of the system which is  a superposition of Excited-atom/Live-cat  with   Decayed-atom/Dead-cat.  The issue is  how to make sense of this situation.  The  proponents of decoherence or decoherence  plus multiple world interpretation offered the  following solution  to the problem:

   Concretely:
 
      \begin{itemize} 
    \medskip
    
    \item       1)  Let us say, we have prepared the system in  an  initial set up, of the box with  the cat   and with 
      the excited  atom,  represented  by the the  state $\psi_0$, and that,
      
         \medskip
         
        \item   2) after a short time, this system  has  evolved into a  typical  state of the form : Dead-Alive cat, that 
      is represented by  the  wave function
        $\psi= c_1\psi_1 + c_2\psi_2$  (where, as usual,  $\psi_1$  represents  the state  of the whole system where the cat 
        is  alive  and $\psi_2$ represents  the state  of the whole system where the cat is  dead). 
      
      \medskip
      
      \item     3) Then, one  invokes decoherence (of any kind), and ends up with a diagonal  density  matrix 
     operator   providing the  description of the system:
     
      \qquad    \qquad  \qquad  \qquad   $ \hat\rho= 
        c_1^2 |\psi_1\rangle \langle \psi_1|  +c_2^2 |\psi_2\rangle \langle \psi_2|  $
   
         \medskip
         
        \item   4) Then,   one  argues that the problem is solved.
      
          \medskip
                \end{itemize} 
        {\bf  This is  the decoherence solution to the problem}. 
      
          \medskip
      
       One might  want to  accept  this  as reasonable.  However    upon closer inspection, it seems rather unclear in what sense does this  mean that the problem is solved.  
      If we take the view that the quantum state represents the reality ``out there",  as would be necessary in  
      any realistic interpretation of Quantum Theory (rather than, say, the view     {\bf b) } of section III), we  would 
       still face a  serious  interpretative issue,  simply because the equation in  item 3 above,  continues to represent a
 situation in which the two alternatives Dead and Alive cat   coexist  (even if they do not interfere).
       
         At this point, we need to rely on an interpretational instruction, for which there  appears to be none else  but  one of the following two OPTIONS:
         
                   \medskip
             \begin{itemize}             
          \item {\bf 5a)} ``One of the two ALTERNATIVES becomes reality, and the other  disappears".  This 
            is, of course,   analogous   to the collapse  (or  reduction postulate). So we are almost back to square one.
          
             \medskip

        \item  {\bf 5b) } ``The two alternatives coexist  in a reality   with multiple  branches. Only  one  of these  branches  corresponds to what we      
          perceive, ( presumably because  we or our mind is entangled with the system).
            This seems  to  be the point of view adopted, in general, by  a big  part of the school  advocating the decoherence type of 
          solutions to the measurement problem  and it seems  closely connected with the "many worlds"  
          interpretation. 
          (In fact this   would be the the view offered  by  the  `` decoherence plus  many worlds" interpretation).
              \end{itemize}             
       \medskip
                    
     Thus  the position represented  by  {\bf 5a)} does not help,    while that of   {\bf 5b) }  has   a   serious problem, which  has  to  do  with the issue of  how  do we treat the ``things that happen"?
    
      \medskip
      
     The  problem in this particular case  is the following:
               The posture {\bf 5b)} indicates that we only perceive one branch of reality,  and that  there 
                are many other branches ``out there",  about  which in  most cases  we can know nothing at all.  We  would know  of a branch  that  is not ours  if we are aware that 
                a particular  current condition of our world   follows  from a  previous   situation  which  was somehow a   superposition  of   the current one  and  some  others  that must have branched  out.
               However most of the time  we would simply not know anything  about these other coexisting branches. Moreover,  it  is  clear  that  if  we take such view,  we  must accept 
                 (unless re-coherence is  forbidden in principle within some   modified version of the theory) 
                 that some  of  those  other branches,   might possibly re-cohere with our own branch, leading to completely unpredictable, and yet  in principle, observable, effects. 
                 We  would not be  able  to justifiably argue that re-coherence is  impossible, or  even unlikely, in any concrete situation,  simply  because,   by hypothesis,   we do not know  
                 what is going  on in those  branches  that we do not perceive.  In fact, this posture  seems to eliminate 
                  any justification  we might have  in stating that the  initial (would be)  wave function in  the previous
                   example, corresponds to that of  {\bf 2)}.  This is so, simply because  we could not, at step {\bf 1)} be  able to   argue that
                     we had described the totality of the (relevant) existing reality, any more than we could,  in  
                     step {\bf 5b)} argue that,  say,  the Dead  Cat alternative, represents the totality of  the 
                     (relevant)existing reality \cite{B. Kay}.  
                     That is,  if we took this  approach, we would  be  forced to acknowledge that  in exactly  the same fashion that the
 Live Cat alternative, is part of the wave-function representing (together with  the Dead Cat  alternative),  the full situation at  stage {\bf 5b)},  
 we  would have to accept that,  in principle, there  exists   out-there many  other branches   or  alternative realities  that should be incorporated in the description of the  full situation at  stage {\bf 2)}.

Thus,  the decoherence ideas,  even  if   taken  together with the many worlds  interpretation,  clearly fail to  offer a satisfactory resolution of the matter in  general \cite{MWI}, and, in particular,  
it fails   to do so in connection for the situation we face here\footnote{The problem of the quantum  origin of the  cosmological structure  has  some degree  of overlap  with the  Quantum Suicide  contemplated  
  in \cite{QSuicide}, and, in fact,  is, if  anything,  even more dramatic and relevant than the latter. }.

  \medskip
     
\section{ The exacerbated problem: applying quantum physics to the early universe}

  We should point out that some  researchers in the field, such as \cite{Padmanabhan},  have acknowledged that there is something  mysterious in the standard account  of the  emergence of structure,  and people  like J. Hartle \cite{Hartle1}  have pointed out the need to generalize quantum mechanics to deal  with cosmology, and, of course,  R. Penrose, who in his last book  \cite{RoadToReality} has stressed  the relevance of the general measurement problem in quantum mechanics to the problem of breakdown of the H.\&I.  during inflation,  comparing  it with the  problem of the breakdown of spherical symmetry in a particle decay.
In our  view, this  analogy  does not emphasize the point that, in the cosmological context,  the problem is  even more severe than in ordinary situations, because, in that  case,  we can not even rely on the  strict  Copenhagen interpretation  as a source of  ``safe and  practical rules".  We  will come  back to the  specific  example  in Section VII. 

  The  exacerbation  can  be illustrated  with   an example that exhibits quite  clearly  the deepening of the problem in this context: Let us consider the following  quotation from  a  well known thinker on these   issues in quantum theory:
   {\it  ``As  long as we remain within the realm of mere predictions concerning what we  shall observe (i.e. what will appear to us) and refrain from stating anything concerning ``things as they must be before we observe them `` no break in the linearity of quantum dynamics is necessary ".}  D'Espagnat 
   \cite{DEspagnat}.
    
{\it \bf However, in the cosmological setting, we need  to deal, precisely,  with this situation:  we need to think about the state of affairs of the universe before the emergence of the conditions that make us possible, before we existed and before we ever carried out an observation or measurement. }

 That is,   we can not rely on ourselves  and our measuring apparatuses, as part of the explanation, precisely because both  are  the result of the  emergence  of the conditions we want to explain.

 Other  avenues  that might  be  used  in different contexts  to address the problem,   even partially; are equally closed in this case:  Inflation is   supposed to drive all  fields  to their corresponding   Bunch Davies   vacuum, so there is  no  nontrivial environment\footnote{The requirement   that  the environment  be nontrivial is  implicit in the assumptions  about energy exchange and information fluxes  from the system  to and from the the  environment and can bee seen  explicitly for instance in \cite{Zeh} and  similar treatments of the  emergence of classicality and related  issues.}, and there are evidently  no  external measuring apparatus.  In short, in  the situation at hand,  the problem  is  exacerbated by:  i) the  absence of  a nontrivial   environment,  ii) the  absence  of  measuring apparatus iii) the  absence of observers\footnote{The  fact  that  the emergence of each of  these   components is, in this case, contingent on what we must understand,  indicates that  these   aspects   can not be used   in the  discussion unless one is willing to  accept   some  sorts of circular reasoning.  Needless is to say  that we  believe  scientific  explanations  should  not  allow that.}. That  items i), ii), and  iii)  above,  play a  fundamental role in the standard  arguments for the emergence of classicality,  can  be seen, for instance in \cite{Zeh}.

 Next, we turn to  consider a  particular account  which, as  we  already indicted,   we consider representative of  a  widespread  view held  in the   field of inflationary cosmology, and   which  can  be characterized by the idea that   there is no problem  with the inflationary account  of the quantum origin  of cosmic  structure.  The  objective here is to illustrate in the  concrete  example offered in  that  account,  the  serious problems  that, often ignored, lie  behind  those views.
 
 \section{ A concrete  example}
 
    With all of our  general discussion  in mind,   we  will now look carefully at one of  the most  explicit  
     of the   traditional accounts on the matter,  the one
     recently presented  in\cite{Polarski-Kiefer} concerning  the  question:  ``Why  do the
Cosmological  Perturbations Look Classical to us?".
 One further advantage of  using this manuscript  as illustration is that   when the article was written,   the authors  were aware of the generic  criticisms made  against the standard  accounts  in \cite{Us}, and elsewhere.
 We should stress that  many of the remarks that lie ahead have been  discussed, albeit in a  less specific setting,  in \cite{Us}, with some 
  other points  having been made indirectly  in \cite{Anandan}. When  following this section of the manuscript it is convenient for the
   reader to have a copy of \cite{Polarski-Kiefer}  at hand. 
  We believe that most of the criticisms  presented here,  apply  equally to many of the works  in \cite{Cosmologists},  
 with  all of them  affected  by  at least one  of the generic  problems we have  reviewed. 
  However, the  detailed discussion  of  each one of those  works   and their  specific  problems  is  well beyond the 
  scope of this paper.   Nevertheless, it is not hard to see, the source  for the  generic shortcoming, which is   simple and  very well known:  
   Given  a system  whose initial   quantum state  has  a given symmetry (in the  situation at  hand the symmetry is homogeneity and isotropy), there  is no  mechanism by which 
    the  standard unitary  evolution  governing the system,  would result   in a state  lacking  that   symmetry,  as long as   the dynamics 
  governing the evolution respects  the symmetry. To  escape  
  this well known   fact  about  quantum theory, the  various proposals   have to resort to 
some kind of  approximate description of  a partial system,   and assigning to  that partial description  a superior  value  overriding the 
 precise and generic  conclusions mentioned  before.  The point is  that  in the cosmological setting  we 
are dealing with  here,  there is no justification for doing so.  In fact, when considering any explanation 
where the conclusions  from the  {\it approximate analysis} (such  as those of \cite{Polarski-Kiefer, Cosmologists}),  
come  into conflict  with the {\it rigorous and general results}, such as the ones  we have just  expressed  above, 
it is  clear that the  approximate  analysis  must be flawed  (the flaw  often residing in  the implicit introduction of unwarranted  assumptions).

 In  considering  in detail  a discussion  such as that  in\cite{Polarski-Kiefer},  the first thing that one  should  consider,  is whether or not   the  approach  described  in  that manuscript starts out  by taking
an exceedingly  narrow view of the issue.  In fact, it is  not  uncommon to encounter  analysis  where   the underlying assumption  is  that all what needs to be understood is  ``why  do we see the perturbations as classical?'',  while taking as  {\bf a given that we are here} with our machines and their limitations, to  act as observers of the cosmological perturbations\footnote{In fact,  when we consider the question posed as  ``why  do we see the perturbations as classical'' one can not but wonder, what should we have in mind when considering   the alternatives  that such question seems to evoke.  In other words,  what would it mean for us, not to  see the perturbations as classical?  What would they look like if we saw them as quantum mechanical?
Clearly bringing our own selves into the center of the question in this  way, transforms the issue into  something that can not  be addressed in the  simple terms that  the approach  of the manuscript  is  based  on.}, and {\it ignoring the fact that we are one of the outcomes of these perturbations}.  This  view of the problem seems to  ignore  that the issue  is set  within the more general problem of cosmology. The point, as we have already noted,  is that  cosmology is the quest for the   understanding   of how the universe evolved, 
including 
the conditions that  resulted  in late appearance of human beings  capable of looking at the sky  and of  studying the traces of the perturbations in question. 
Therefore, by limiting the question  to an issue  of observation,  one  would  be implicitly putting into the formulation of the problem  what should indeed be part of the answer:  That the universe   eventually develops the structure  that makes feasible the  emergence of human beings  capable of viewing the CMB  sky. Once that  fundamental  aspect of the problem is overlooked, the  question becomes simply  why the  spectrum has certain detailed  features, and  a big part  of the main problem is bypassed.  In fact,  such  approach is,  in a sense,  the application to  quantum problems in  cosmology, of the  generic approach to the measurement problem  in quantum mechanics
that starts off by assuming that one of the parts of the subsystem is  a classical  apparatus, or that there is   a classical observer,   whose  origin  and description is not attempted,  and which provides the set up with a clearly identified  referent  for the use of the Copenhagen interpretation or something very  close to it. This is clearly a  rather  unsatisfactory   way to address the issues of cosmology  from within a quantum theory of reality, unless, of course one takes the view that quantum theory is not such a thing,  but rather one of the unpalatable possibilities enumerated in section III.  Any such approach, as indicated in  the previous  section,  would  call into question the whole  justification  for applying  quantum theory to cosmology.

 Furthermore, apart from the  very  general  issues above, we will see that  the   general arguments    behind 
  the  conventional accounts on the question 
   rely on, what   a detailed examination  shows are,  inappropriate interpretations and  problematic arguments.   In order to see  these aspects in detail,  one must  look at an  individual proposal in some detail. We  will  focus  on  the  above  mentioned proposal for  concreteness. 
  
The  setting is  that of a background  Robertson Walker  space-time undergoing
exponential inflation and described with the metric:
\be
    ds^2 = a(\eta)^2\left[
  -d\eta^2 +  \delta_{ij}\, dx^i dx^j
\right],
\ee
  where the scale factor is given by  
  $a(\eta) = -\frac{1}{H_I\eta}$, with $H_I^2 \simeq
(8\pi/3)GV$ with $V$,   the inflaton  potential,   and  the field $\phi_0$ in slow-roll
regime.  The conformal time  $ \eta $ running from $-\infty $ to $0$.   
The  quantum aspect of the scalar field $\delta\phi  \equiv \phi - \phi_0 $ is
  described in terms of a rescaled field $y\equiv a\delta\phi$.
  The field is written in terms of creation and
   annihilation operators and written in terms of  its  (spatial) Fourier transform:
  \be
   y(\eta,  k)= f_k(\eta) \hat a_k +f_k^+(\eta) \hat a^*_{-k}
  \ee  
 with $ f_k = \frac{-i}{\sqrt{2k}} e^{-ik\eta}( 1-\frac{i}{k\eta})$
 and   where  the conjugate momentum 
   \be
   p(\eta,  k)= g_k(\eta) \hat a_k +g_k^+(\eta) \hat a^*_{-k}
  \ee  
   with $  g_k =  -i\sqrt{\frac{k}{2}} e^{-ik\eta}$
   (Note  that we have suppressed the distinction of  the 
   vector $\vec k$ and its magnitude $k$  for simplicity of notation).
    
     The approach  followed  makes use of the fact one can write
   these operators as combinations of  the following  time independent  operators:
   \be
   \hat Y( k)=  \frac{1}{\sqrt{k}} (\hat a_k + \hat a^*_{-k}),  \qquad \& \qquad
   \hat P( k)=\sqrt{\frac{2}{k}} (\hat a_k - \hat a^*_{-k}).
     \ee   
where the  coefficients  are:  
\be
f_{k1} = \Re (f_k ) = \frac{-1}{\sqrt{2k}} [\sin (\eta k) + \frac{1}{\eta k} \cos((\eta k) ], \qquad 
f_{k2} = \Im (f_k )= \frac{-1}{\sqrt{2k}} [\cos (\eta k) - \frac{1}{\eta k} \sin((\eta k) ],   \ee
and
\be
g_{k1} = \Re (g_k ) =  \frac{-1}{\sqrt{2k}} \sin (\eta k) , and \qquad  g_{k2} = \Im (g_k ) =\frac{-1}{\sqrt{2k}} \cos(\eta k) \ee
 respectively.  
 That is 
   \be  \label{PK24}
   y(\eta,  k)=\sqrt{2k} f_{k1} ( \eta) Y( k) -\sqrt{\frac{2 }{ k}} f_{k2} ( \eta) P( k),
  \ee   
 and 
    \be \label{PK25}
   p(\eta,  k)= \sqrt{2k} g_{k2} ( \eta) Y( k) -\sqrt{\frac{2 }{ k}} g_{k1} ( \eta) P( k).
     \ee 

The  first set of considerations  offered in  \cite{Polarski-Kiefer}, in the section "Quantum to Classical Transition: the pragmatic view ", are based on the properties of these functions.
 The point is that in the limit $ k \eta \to 0 $  (corresponding,  for a given $k$,  to the infinity future of a never-ending inflationary era)  the coefficients $f_{k2} $ and  $g_{k1} $  vanish.     
      \begin{itemize} 
             \item
             This
 is next used to argue that  { \it the non-commutativity of the operators  $Y(k)$ and   $P( k)$ becomes irrelevant}  and can be ignored as 
 it concerns the quantum field and its momentum conjugate. 
 \end{itemize}
  
  The first thing we note is that this discussion, so far, applies to the quantum  field operator itself, and, 
  thus, nothing of what has been said  up to this point is limited to the vacuum state alone. Thus, in principle, 
   it should apply to all states. The quantum field has become equivalent in all generality  to a  classical  field.  
   But why should this be limited to the inflaton field? All fields present during  inflation should be subject to the same analysis. 
   Thus,  we would conclude that all  fields  in  nature are equivalent  to their classical counterpart\footnote{That is, unless one  wants to argue somehow  that other fields,  say those of the standard model of particle  physics,  were not  affected  by the inflationary  expansion  which  drove the  inflaton to the  vacuum state.  Such idea  would lead  us  to serious  conflicts with the basic  
   tenant  of General Relativity:  the equivalence principle,  and  thus  we  will not  further consider that problematic  possibility here.}. After all, every field that exists 
   should be present in its  corresponding vacuum state during the inflationary epoch. The fields are not  simply turned on when we  need
    them. This is required  by the working hypothesis behind all of physics  regarding the {\it invariance of the laws of physics in ``time"}. 
    The conclusion is, of course, incorrect for other fields,  why should it be correct for the inflaton? It seems that
     the only possible posture here is that the modes of all quantum  fields have  become classical  if they  correspond to wavelengths that are ``large  enough".  But then we must ask ourselves:  large enough in comparison to what?   the human scale?   perhaps  the current horizon scale?  Anyway,  it seems clear, that in order to justify this approach, one would need to introduce several other {\it ad hoc} rules into the theory.  Moreover, as it has been recently 
      discussed in \cite{HollandsWald},  the position that one could consider the quantum field simply  as a collection of independent modes, as would be needed in order to subject the different modes to very different treatments (such as considering some  to be classical, while some others are considered as fully quantum mechanical), {\bf is untenable} in  light of the  locality of  the field theory  and the need  for composite operator renormalization.

    The next thing we should note is that,  when speaking about an operator, the issue of what is small and what is not can 
    only make sense  once one has identified the way one wants to extract a number from the operator at hand.  In general, one would need  to consider the
      matrix element of the operator in  a certain class  of states,  including the states  constructed  through the application of  other relevant  operators on
        certain collection of relevant states (in our case,  that would involve the states constructed out of the vacuum by suitable application of  field  operators, i.e., the Fock space).
         Moreover, it is not at all clear why one should not consider  one of the  most direct, the c-number«s  that can be extracted 
          from the operators at hand:  the  commutator  between the field   and its conjugate-momentum. Taking this more 
          rigorous mathematical stand,  one sees  that it is  rather unclear how  can one convincingly  argue that the 
          operator  in question vanishes. In fact, the decoherence arguments of this type 
     rely up to this  point, on the behavior of the field and momentum conjugate as  operators, and not just on particulars of the state. Thus, in accepting them,  
     we would  have to conclude it is not just  the vacuum state that becomes  classical,  but the full  quantum mechanical  system, as a physical entity.
     And  if this is the case for this  field, we  must in turn  question,  why  would it not  be the same  for  the other fields?
      
    Finally, we should recall  that  inflation is not supposed to continue indefinitely, but  must stop at some 
      point (even if  something like 80 e-folds  of expansion are accomplished while it lasts) the quantity that is being discarded in  
      the arguments  of \cite{Polarski-Kiefer} is not,  strictly  speaking, zero (in any of the senses that can be given to this  statement,  as discussed above).  One  can argue that is   extremely small, and   thus negligible,  but in order to do that one would have to face other issues:
        1) In comparison to what is it to  be taken as small?  certainly not as its contribution to the commutator of field  and momentum conjugate is concerned. 2) Why should  we trace over it?    presumably because the quantities  are to small to be measured,   but this, of course, can only  mean, ``to be measured by us  with  current technology"  as  there is no possibility of  arguing about the un-measurability of those   quantities by other  conceivable  beings, or even by  future generations of  humans.
        
        This reliance on what is  experimentally  accessible to  us  (or that which the author envision as being attainable in the  relatively near  future) is prevalent throughout  the   decoherence type  of analysis. We find this, in  \cite{Polarski-Kiefer}:  for instance,  when  considering the separation of the  field $y$ and  momentum conjugate $\pi $ (equations 24, and 25, of that work, which correspond to equations 
        (\ref{PK24}) and (\ref{PK25}) of  this manuscript ) and the reliance on the existence of modes characterized by the functions $f_{k2}$ and $g_{k1}$ that  ``become vanishingly small" as a result of inflation.  These are thus considered  as unobservable, negligible,  etc.  As we have indicated,  such   reliance, is  unacceptable for a theory that is  supposed  to  explain our own origins.
         
  The  ease with which unjustified assumptions appear, often  unnoticed,  in this kind of   discussions can be further exemplified  by considering the concrete argument in more detail: The first point  is that such functions only serve to connect the operators $ \hat y(k,\eta)$ and  momentum conjugate $ \hat p(k,\eta)$  with the operators    $ \hat Y(k)$ and $ \hat P(k)$.   The point is that,   even if  we identify an operator $\hat O$  which is written in terms of other operators  with coefficients that  vanish in a certain limit , we must   concern ourselves with how is this connected with the quantities we measure.
        In the inflationary setting,  we can always  construct the operator  $ a(\eta)^n \hat O_k(\eta)$  and note that, for an appropriate choice of  $n$,  it does not decrease at all.  Actually,   in the case at hand, we can even look at the commutator  as an observable that does not tend to zero  even if $a \to \infty $. 
       In fact, it is worthwhile mentioning that  if we  consider, instead of the rescaled field $\hat y(\eta,  k)= a \hat \phi (\eta,  k)$   and its conjugate momentum  $ \hat p(\eta,  k) $, the  original un-scaled inflaton field $\hat \phi(\eta,  k)$ and its conjugate momentum $\hat \pi =  a   \hat p(\eta,  k)$,  we  end up replacing the functions $f_k$ and $g_k$ of  section II of \cite{Polarski-Kiefer} by $f_k' = f_k/a$ and  $g_k'= g_k a$,   and  when resorting to a similar analysis, as done in that paper, we  would see that the mode $ g_{k1}$ no longer vanishes in the   $ a\to \infty $  limit .   The ratio of the different  modes, is, of course, unaltered, but in order to argue for the un-observability of  such mode, one  would have  to  argue that the big value of the other mode,  somehow makes this mode unobservable,  and, in order to justify that, one would 
   have to get into the issues regarding the possible precision of instruments, 
    as  an essential part of the arguments. 
 
Furthermore,   such accounts  tend to make  heavy use of unjustified  identifications, that  rely on mixing up classical and quantum arguments:
We  can see this, explicitly, at several points in the example provided  by  \cite{Polarski-Kiefer}: 

\smallskip
    \begin{itemize}      
    \item    
   For instance below equation 26  of  that  manuscript 
 \be
 p(k,\eta) = p_{cl}(k,\eta)=\frac{g_{k2}}{f_{k1}} y(k,\eta)
 \ee
 (where presumably the  assumption is that one should  focus on  something like the value of $p$ corresponding to the maximum value of the Wigner functional,  considered  as a function of $p$, with the  given  value  of $y$  held fixed),  we read: {\it``For a given realization of the perturbation $y(k,\eta)$, the corresponding momentum $ p_{cl}(k,\eta)$ is fixed  and equal to the the classical momentum corresponding to this value of $ y(k,\eta)$. Then the quantum system is effectively equivalent to the classical random system, which is an ensemble of classical trajectories with certain probability associated with each of them".}
      \end{itemize}      

 To  uncover the problem we must   consider the above  statement with   some care: First of all,  what justifies talking about a given realization? What are these realizations?  Where, in the  standard quantum  theory, is there any reference to  ``a realization"?  In fact,  one must wonder if  the position advocated  by followers of these  type of approaches is  that 
each one of those realizations is characterized by a definite value of field and momentum conjugate, as characterized   by equation 26 (reproduced above) of \cite{Polarski-Kiefer}?  Perhaps,  we should  think  that this  only refers to the mean values of  the corresponding quantities  in some quantum states (one  for each  realization)? If so,  which ones? 
Then, the  manuscript states  that  
{\it  ``the quantum system is equivalent to classical random system"}, indicating that, according to such posture,
  there is associated to our quantum system a given  {\it ``probability assigned to each of the elements "} of a certain  classical ensemble. 
 What is this  classical ensemble  supposed to be representing ?  Should  we view  our universe as  an element of an ensemble of  classical universes?  
In that case, our particular universe would be classical!. Perhaps, what the  manuscript refers to,  implies the   association with a  classical ensemble 
 to the collection of possible outcomes of certain measurements. 
That is to say, to an implicit assumption   that the  quantum  state is not characterizing the objective physical conditions of the system under description, but  it is just a codification of the results of subsequent measurements.
This does not seem to have any justification,  unless  one is   taking the view a)  of section III, which, as we have seen, does not allow us to apply quantum theory to cosmology.

 \begin{itemize}    \item  Regarding the  state of the field, the situation at hand  is simply  that, in terms of the original creation and annihilation operators, the state we identified  as a vacuum  is now a squeezed state.  
  But having a  squeezed state has no effect on the quantum field and  momentum operators
  which  do,  of course, continue to satisfy the Heisenberg uncertainties.  {\bf The manuscript seems to ascribe a fundamental value to the squeezing of the state, ignoring that one can always find a new set of operators in terms of which, the  evolved vacuum will look as a ``standard vacuum".}
    One can see exactly what lies behind such arguments by considering 
   a simple harmonic oscillator in Quantum Mechanics.  The Hamiltonian $ H= \frac{p^2}{2m} +  
   \frac{ m\omega^2 x^2}{2} $,  (with $\omega^2= k/m$) and standard commutation relations $ [p,x]= -i $.  We can write the usual creation and annihilation operators as 
   \be
     a= \frac{e^{s_0}}{\sqrt{2}} x +i  \frac{e^{-s_0}}{\sqrt{2}} p, \qquad a^\dagger= \frac{e^{s_0}}{\sqrt{2}} x -i  \frac{e^{-s_0}}{\sqrt{2}} p
   \ee
   with $ e^{s_0}= \sqrt{m\omega}$.
    These operators satisfy the usual commutation relations $[a,a^\dagger] =1$.
   We  can now define (without changing the system or its  hamiltonian), for arbitrary values of  $s$,   new operators:
     \be
     \label{at}
     a_s= \frac{e^s}{\sqrt{2}} x +i  \frac{e^{-s}}{\sqrt{2}} p, \qquad a^\dagger_s
     = \frac{e^s}{\sqrt{2}} x -i  \frac{e^{-s}}{\sqrt{2}} p
   \ee
   which  are related to the original creation  and annihilation operators  through a ``Lorentzian rotation". 
   In fact, these operators satisfy the same commutation relations  $[a_s,a^\dagger_s] =1$.
   However,  following the logic of  \cite{Polarski-Kiefer},   we would have to  say,  by looking at equation
   (\ref{at} )that {\it in the limit when  $s$ is very large (i.e. $s\to \infty$) the  fact that $x$ and $p$ do not commute 
    becomes irrelevant and that,  therefore, we are in an essentially classical situation, 
     where for each  value of $a_s$  there is a corresponding value of  
    $a^\dagger_s$}  (which, in this case, would be the same value).
This argument would imply that,  through the simple act of choosing to express things
 in terms of suitable variables,  we can change the nature of a purely quantum mechanical
  system into something which is essentially classical.
  The analogy with 
  the situation being examined, 
   can be  further strengthened  
 by noting that, as a matter of fact, the  above construction is just what is used to define
  squeezed states,  namely, the states annihilated by the operator $a_s$  are squeezed states  
  when seen from the point of view of the usual basis  of the harmonic oscillator Hilbert space:  
  $\lbrace |n \rangle ^{osc} \rbrace$. Analogously, the  usual ground state of the harmonic oscillator would appear as a squeezed state  when viewed 
 in the basis of generated by  the repeated application of $ a^\dagger_s $ on  the ``displaced vacuum"
  $|0,s>$ ( the state defined by   $a_s|0,s\rangle  =0$ ).
  Therefore, the claim made just below  equation 26 of  \cite{Polarski-Kiefer},  that  such equation 
``is true for the quantum system
(in the operator sense)" is incorrect.
  If we try to consider  how  these  lines of argument  might be justified,   
  we would have to assume that there  is  an implicit assumption  about a  notion of a 
  ``preferential role"  for the particular operators  that   represent variables that are ``more natural" 
   than others,  and that there is some particular aspect of the  state of the field  one is dealing with, 
    that  bypasses the counter-arguments  we have exposed above (some  sort of argument that would  say,  for instance,  that  the relevant parameter
      in this case is  $s_0$  rather than the ``less natural" parameter  $s$).  That  seems to be the tenor of part of the discussion  in section V of \cite{Polarski-Kiefer},  but  we will get to that shortly  and  will see that it also  fails to pass  further scrutiny.
However,  before doing so,  we must lastly point out that a related  situation where  $s$ is  very large  can  be indeed achieved by forcing 
the ``natural" parameter $s_0$ to  be very large,  corresponding, for instance, to  the situation where $m$ is very large  for fixed value $k$. Actually,  we can choose both, $m$  and $k$ to be very  large,  while keeping the value of $\omega$ fixed at any desired value.  But it should be clear that,   in that case,  the ground  state of an  harmonic  oscillator is not a  classical state. For instance, the uncertainty of the  momentum  is much  larger than the corresponding expectation  value, and  the  energy levels of  the system   are separated by the usual $\Delta E = \hbar \omega$.
Finally, we must note  that these squeezed states of the electromagnetic field\footnote{Defined  with respect to the usual construction 
 of Quantum Filed theory in Minkowski space-time.} are well studied  in the field of quantum optics,  and, in fact, rather than being ``almost classical"  as indicated in \cite{Polarski-Kiefer}, they are  known to exhibit  behavior   that  is  quantum  ``in extreme" \cite{Squeezed}. For example,  we note that when a  system consisting of two harmonic oscillators, is  in a state where both degrees of freedom are in squeezed states, the appearance of highly nonclassical  EPR correlations is inescapable\cite{correlations -squeezed}. 
  Thus, in the case of an infinite set of harmonic oscillators  in squeezed states, such as the vacuum  of  quantum field we are interested on,  the   situation  can not be said to be anything but of a  highly  quantum-mechanical nature.
     \end{itemize}     

On further examination,  we find  other problematic  arguments  which are of common usage, being  explicitly  stated  in this section of  \cite{Polarski-Kiefer}:  
 
    \begin{itemize}      
    \item             Below  equation 26 of \cite{Polarski-Kiefer}, we find the following statement referring to  the issue at hand: {\it``an analogous situation happens for a  free non-relativistic particle possessing 
 an initial Gaussian minimal uncertainty wave function....  At very late times the position of the particle 
 does not longer depend  on its initial wave position $ x(t)\sim (p_0/m) t$. We get an  equivalence with
  an ensemble of classical particles obeying this  where $p_0$ is a random variable  with probability $P(p_0)= |\Psi|^2 (p_0)$".}
  Thus, if one  were to accept   such  arguments  without  caveats,  we  would be forced to conclude that,   if we wait long enough, a  free quantum particle  would become  classical.
 One thus, must   assume that  authors  have something  more restrictive in mind.   Perhaps  what they  want to point to,  is   the fact that if  one  were to  consider  the measurement of  the position of  an ensemble  of such particles, a long time   after it was  prepared in the  minimal wave packet,  and   concentrates on distances that are not to big,  the results   of a  series of  identical experiments, would,  --  according to  quantum  mechanics  { \bf including   the reduction  and  Bohr   probability rule postulate}-- coincide  with  those of  a  classical  ensemble they  describe.  However, it  is clear that the  statement would be valid  only if one  has in mind such  ``measurement context", and,  in fact,   a particular  kind  of  observable (such as position)  associated with  such  measurement.  Regarding other possible observables the  statement would  be false. Thus, the  problem with this  sort of   claim,  indicating   that  the situation is   analogous  with that  faced in inflationary  cosmology,    is that  in the latter case  there seems to  be  no identifiable reasonable  measurement context, measuring apparatus, or observable, and  of course  the  situation is  such   that  one  would not  want to refer,  explicitly, to the fact that they are relying on the  uncomfortable  part of the postulates of quantum theory.  The  point  is  then, that statements such as the one in italics  above,  are nonsensical  unless  an appropriate context  is specified, and,  in the  cosmological setting we are interested,   one can not  justify the  identification of any  appropriate  ``measurement context".  The  fact that  such  caveat  is not  mentioned in connection with the  statement in italics above,  a fact  which renders it indefensible,  hides the essence of the problem in the cosmological context.   
      \end{itemize}             

  Next,  we consider the  more refined  arguments  such as those  presented  (on section V  of \cite{Polarski-Kiefer}) under the title 
   Quantum To Classical Transition: Decoherence. There  we see the  recasting  of   the previous arguments  
   relying now, explicitly, on the standard decoherence argument  that  interprets  the fact that  the reduced  density matrix has the  form  used in describing a statistical ensemble, to argue that the system has become one of the latter.  Thus, although no  specific  claim is  made  to that end,  the whole  argument in this section seems   based on   the assumption that decoherence  does in fact offer a   solution to the measurement problem in QM. This is an assumption that, as we discussed in the previous sections,  is not  correct.  
   \medskip

 We  will now look exactly  how and where do the problems arise in this specific treatment.  
  
   \begin{itemize}     \item
   The first issue that is  addressed is that of the identification of the environment:  Here the   analysis  offers  a path  to an answer based on the  notion   that {\it ``.. in any fundamental theory... there is an abundance of different fields with different interactions. Among them  there will  not  be difficult to find appropriate candidates for environmental fields generating decoherence for primordial fluctuations".}
    The problem with this  assertion is that, in the same way that inflation is supposed to drive the inflaton
  perturbations into the vacuum state,  it will drive all other fields  into their corresponding vacuum state.
  In fact,  even if  we accept  that there are interactions  among the various  fields,  the point is that  inflation drives   the whole quantum system of interacting fields into the appropriate  vacuum  of the combined system. 
The equivalence principle indicates  that all fields  will  be affected by the expansion  associated  with the  gravitational  sector, in the same fashion, and, thus,  it would  be  quite difficult
circumvent the conclusion.
  
The  argument  
 then indicates that, even if  appropriate  environmental fields  do not exist,  the fact that the theory is nonlinear indicates that there will be interactions  among the different modes. 
  But again, this   argument seems to ignore the fact,  that, if one wants to take into account such interactions, one must recognize that the vacuum to which inflation drives the fields is the full non-perturbative  vacuum.  One  can not claim that the initial  state of the inflaton is the free vacuum, and then  argue that the interaction plays an important role in generating the desired  classicality.
  Nature  does not rely on perturbation theory,  only our description of it  does.  Moreover,  if one   wants to take into account the self interactions,  the fact that the Lagrangian that describes them  is translationally and rotationally invariant,  ensures that the resulting state  will  be as  homogeneous  and isotropic, (as well as pure) as the original one. This  consideration  thus  breaks  the analogy  with the case of a quantum particle  interacting  with a  bath of photons. 

Then,  it  is  argued,  that the entanglement of the state,  when considering the individual  modes,  is analogous  to  the Hawking and Unruh  effects. 
 One  must  be careful with  taking   such analogies  as appropriate, 
  because, in both of these 
latter situations,  one  is interested in looking at the state,  as it is perceived by observers limited to specific regions of space time,  and  who do not have - by definition- access to  certain degrees of freedom.  In the cosmological case, the manuscript seems to be  implicitly assuming,  that  one  is again interested in   the observations  made by observers  that do not have access to certain  degrees of freedom, but  as we have  argued before,  that is not the case: {\bf In cosmology, we want to understand the conditions that lead to the emergence of such observers (i.e., Us). }
  
   Otherwise,  cosmology should no longer be regarded  as the search for an explanation of the evolution of the universe,   in terms of  the laws physics, but  rather as the  mere attempt at  retrodiction of information,  given the fact that  we are here,  with our particular specificities, including those that limit our ability to monitor  certain degrees of freedom. {\it Why  should   one contemplate inflation  at all,  if that  were  the case?}
    \end{itemize}  
    
    In the  analysis presented  in  \cite{Polarski-Kiefer},  one finds the acknowledgment that the coincidence of certain classical and  quantum expectation values  in a pure quantum  state  is not the same as an ensemble of stochastically distributed  classical values,  but,  nonetheless,  that account  reiterates that a quantum to classical transition occurs  thanks to decoherence and squeezing.
   
    \begin{itemize}  
    \item It  is  argued that the cosmological fluctuations  represent the system (or part thereof ) to be decohered,  while  the  other fields (or the inaccessible parts of the the fluctuations) represent the environment.  As we have  already pointed out, 
     this posture,  leads us to a  circular explanation:  We are the result of the evolution of the initial  cosmic  fluctuations,  and  the emergence of those  is understood in terms of some of our own characteristics  as sapient  beings.
      \end{itemize}  
    This  does  not seems  to be a satisfying view of the explanatory  powers of  cosmology as a science.
  Nonetheless, such   descriptions are  profusely found   in the literature.    
  
  \begin{itemize}
    \item For instance,  in the case  of  \cite{Polarski-Kiefer} (below  equation 47  of  that  manuscript) we can explicitly read:  {\it `` ...it (the matrix density) assumes the form of an approximate ensemble for the various states which occur with probability..."}.
    \end{itemize}  
    
 We note  that,  this is, precisely,  the misinterpretation referred to,  by M. Schlosshauer,  in  the quote  we reproduced in section
  IV.
   
   Again,  some of the most conspicuous  and  clear contradictions   that arise, when  considering   arguments  that rely on the  use  unjustified interpretational  extrapolations such as those  above, can be illustrated   by considering the   
 standard Bohm-EPR \cite{EPR} setup while entertaining the notion that the two particles might have a particular spin  orientation  in the absence (or before) a  measurement is carried out \cite{Mermin}.
 We could, for instance, decide to trace over  the spin D.O.F. of one of the  particles of the EPR pair and obtain, for the spin  D.O.F. of the other  particle, a diagonal density matrix, and be tempted to interpret this as indicating that the particle has one of the two spins orientations.  But we know this leads to contradictions:  It allows one to deduce that the Bell's inequalities should hold,  but these are  not only contrary to Quantum Mechanics, but have been shown to be violated experimentally \cite{Aspect}.  Such position is thus untenable.
  
Besides these,  one  often can  find  additional problematic  aspects of in  specific  the traditional  treatments on the  subject. Focussing again on in the  account of \cite{Polarski-Kiefer},  one can  identified   the following  aspects.
    
First is the arguments  on which these type  of approaches  must  relay  very heavily when  making the   selection  of  the preferred   basis of the Hilbert  space,  that of the so called ``pointer  basis".
      
    \begin{itemize}      
    \item       Regarding this issue  we find (below  equation 48 of  \cite{Polarski-Kiefer}): {\it  ``fields coupling to cosmological fluctuations  are expected to  couple to  
     field amplitudes  and not canonically conjugate momenta of field amplitudes"}
     One  should  question why  would this be  so?  
    For instance, should one not expect  there be  vector fields $ A^{\mu}$, say,  similar to gauge fields, that couple precisely to the  combinations of the form $\psi \nabla_\mu  \psi$?  Doesn't the  reheating process require a coupling (even if indirect)  of the inflaton with all ordinary fields,  including the electromagnetic one?
    Why can't  the inflaton field  be a part of a multiplet coupling to  a non-abelian gauge fields? 
    What   about couplings to gravitation which should be universal  and then include all forms of energy momentum including that associated  with the  conjugate momenta to the field amplitudes?   Perhaps the argument would be  that the coupling to gravitation is too small and thus the correlations with the metric variables  is negligible,  but then,  how  do  we make sense of the fact that it is precisely through the imprint on the metric and the  corresponding  gravitational red-shift that the Newtonian potential induces in the photons, that we do observe the CMB anisotropies?
      One should further note that  the postures in this  regard,   such as that  taken in \cite{Polarski-Kiefer} go against  the usual understanding of when a situation is classical:  Following the point of view  advocated  in  that manuscript, those  squeezed states our colleagues quantum optics work so  hard  to construct,  would have to be regarded as classical  states. In fact,  we should  recall that one  normally considers  that  a situation can be described as ``essentially classical" when the  quantum uncertainties  are very small in comparison with the  expectation values of the  quantity of interest,   or in comparison with the precision of our analysis.  
       \end{itemize}     
         
          The attempts  to  deal  with the  general  question facing  us, leads  many researchers to twist  the usual  understandings and interpretations  of our physical theories  in  so many ways,   that it often leads to  postures that  are  actually self-contradictory. 
          
            \begin{itemize} 
             \item
            For instance,  in  section IV of   \cite{Polarski-Kiefer}, after  having  argued  that the situation after inflation is  essentially classical, the manuscript instructs us to  rely,  for the relevant predictions, on the quantity $  \langle 0| \delta \phi^2 |0\rangle $ (eq.  (42) of  \cite{Polarski-Kiefer} ), which is nothing but the uncertainty associated with the quantity $\delta \phi$  whose expectation value is ZERO.   In fact, we are  warned that in computing the power spectrum, what we must compute is the ``quantum average".  That is,  the instruction is that  one  should not rely too much on all the arguments indicating the situation was classical. It seems  that in  the end one  does  not know,  if  the situation is essentially  classical or is it not.  There seems to be  an internal   contradiction  when the  procedure calls   for the  evaluation of   the relevant    observable  quantities  using a quantum object,   if the situation  looks classical.  In fact,  one  must  wonder,  why should we base our  calculations  on the vacuum state rather than on one of the states  $ |n \rangle $ in the ensemble of states corresponding to:
            \be
          \rho_s   \approx  \Sigma_n |c_n|^2  |n \rangle \langle n|
          \ee
       which  according to equation  (47) of  \cite{Polarski-Kiefer}, is supposed  to emerge from the decoherence analysis\footnote{In fact, we could now turn  around and  ask for instance: what exactly are these states $|n \rangle $?  
 What are, for instance, the uncertainties  in the  field and momentum variables for each of these states?  How would we even proceed in attempting to answer such questions?}? Aren't  these, according to such approaches, 
  precisely  those  states  that look classical and which we are  supposed to observe\footnote{It is actually, worthwhile  noting  that  most of the ``standard"  accounts indicate that we that we  perceive  is some sort of ``inhomogeneous   and anisotropic component  of the full quantum state"  but are  conspicuously reluctant to  characterize such  components,  at the quantum level, any further than that, despite the fact that these are precisely the states that lead to  what  we  are supposed to observe.}?
           \end{itemize}  
           
          It thus seems, that  in order to  subscribe  to   the  posture  adopted in those  proposals, one must
         take it  to mean  something like  ``The cosmological fluctuations look classical to us  but in reality the state of our  universe is the  result of the unitary quantum mechanical evolution of the  vacuum, and it is thus still homogeneous and isotropic. It just does not look like that to us". This, in turn, takes us to the posture discussed in 5b)  which, as we argued,  would make it impossible to use quantum mechanics at all. 

Therefore, we must conclude that  the position  taken in  accounts  such as  that of \cite{Polarski-Kiefer} is self contradictory:  We  are supposed to observe the universe in  one of the states -- one of those schematically shown   in the equation above (corresponding for example  to equation  47 of  \cite{Polarski-Kiefer})-- that make up the superposition  in which the system actually is,  but on the other hand, in order   to compare with  observations  we must make use of the  full superposition, i.e.,  the vacuum state.  Is it not the view advocated by \cite{Polarski-Kiefer} that we live in one branch and thus only see the  corresponding state\footnote{It is worthwhile mentioning that one would need to take into account  the evolution of the  expectation values of field operators in the corresponding state, if, in the analysis of the problem,  one chooses to follow such posture to its full conclusion. This would, in turn, lead to the uncovering of the type of  unconventional effects  analyzed  in
  \cite{Us, Adolfo}.  } $|n>$?. If  so, why should  the comparisons  with observations  be made using the full fledged superposition associated with the vacuum state?   A  self consistent posture on the matter at  hand  would prevent the  simultaneous  advocacy of the two  conflicting  points of view.

   \section{Other  Ideas  for a Resolution of  the  Problem   with Standard  Physics}
   
  There are some  other lines of thought which, although  are much less popular, have been considered  as offering a resolution of the problem at hand,   while relying on established physics.   
  As it turns  out,  a closer inspection of such ``resolutions"  shows  that they  are based on inappropriate analogies.
    For instance, one often  hears  that the problem  is completely analogous  to the  decay of an excited  atom or nucleus,  from a spherical  symmetric  state to  an unexcited nucleus or atom  and  a particle that  is  escaping  along a  particular  direction,  which  is clearly not a  spherically symmetric  state.
    
    The  problem  was considered in \cite{Mott},  in  early days of quantum theory, and  its  treatment is thought,  by  many colleagues, to  have clarified the issue completely.  But let us  look at it anew: 
      The setting considered consists of   a nucleus located at the origin of spatial cartesian coordinates ($ \vec X= \vec 0$) in  an  excited (unstable)  state  $|\Psi^+\r$  which  is spherically symmetric,  and ready to decay into an unexcited  nucleus $|\Psi^0\r$,   plus an  $\alpha$ particle in state $|\Xi_\alpha \r $,  which  is  also spherically symmetric.   The setting includes  also two hydrogen  atoms  with their  nuclei  fixed at positions  $ \vec a_1$  and $ \vec a_2$, and   their corresponding  electrons, in the  corresponding ground  states.  
      The  issue   that  is  discussed is the degree to which the nuclei  should be aligned  with the origin (  i.e $ \vec a_2= c \vec a_1$  with $c$  real)  if both  are  to be excited  by the  outgoing $\alpha$ particle.
      
      The analysis indicates  that  the probability of both atoms  getting excited  is significant only  when there  is  a large degree of alignment,  thus explaining the fact that the  $\alpha$  particle   traces  straight paths in a bubble chamber.
    
 Thus,  although,  at first  hearing about this,  one  might think  that  one  has an example  in which  an initial state  possessing spherical symmetry (  $|\Psi^+\r$)  evolves  into  a final state   without such symmetry,   despite the   assumption that the  hamiltonian (governing the decay
 $|\Psi^+\r  \to | \Psi^0\r |\Xi_\alpha \r$   and  the  $ \alpha$ particle  evolution) is  symmetric  under  rotations,  a closer look  reveals  the fallacy:  As indicated, the setting includes the two unexcited atoms, which,  through the localizations of their nuclei, break  the  rotational  symmetry.  Indeed,  the  discussion is  based,  not on the  Hamiltonian  for the  evolution of the  free  $\alpha$ particle,  but rather  on the  Hamiltonian for the joint evolution( including the interaction) of  the  $\alpha$  particle and the  two electrons  corresponding to the two localized  hydrogen atoms.   In fact, the projection postulate associated  with a measurement is  also coming into play in the  analysis of \cite{Mott}  when computing  probabilities  by projecting on the subspace corresponding to the  two atoms  being  excited.  
  It is clear  that  if we were to replace these  atoms  by some hypothetical  detectors whose  quantum description  corresponded  to spherically  symmetric  wave  functions ,  each  one  with support, say,   on  a thin spherical shell with  radius  $ r_i$,  a similar calculation would not lead to straight lines,  but  rather it would in  a spherical   pattern of  excitations. We  would  find that  there was a certain probability of the detectors corresponding  shells  $ i^{th} \&  j^{th}$   being excited  and the symmetry would not  have  been compromised.    The  inflationary situation  we  musty  face is,  in this regard, if anything, closer the the 
  later  rather  than the former version of this  problem.

    Another example  of the reliance on inappropriate  analogies is found in  \cite{Jerome 1} which  proposes  an analogy between the  process that lies at the origin of  the anisotropies and inhomogeneities in the early universe, with the process of particle creation out of the vacuum in the presence of a sufficiently intense electric field, ``the Schwinger process".
    Such type of  analysis  of the Schwinger  process consists  in the evaluation of  the  $S$ matrix element between the "in" vacuum and the "out" vacuum,  and  the identification of the difference between $1$  and   the result as a measure of particle creation. This identification would be  justified  by unitarity, and the observation that the $S$ matrix element between the ``in" vacuum and the other vectors in the ``out" Hilbert space, such as a specific state with  a single  electron-positron  pair, would  be  interpreted as the
 probability of  pair creation in such state. This  would, in turn,   be justified  by the standard quantum mechanics postulate of the projection postulate, indicating that $ | \l A  | B \r |^2$ is the probability  of  {\it finding} the system in the state $|A \r $, if it was originally prepared in  a   state $| B \r$.  However, note  the  need  to rely  on an observation or measurement, implicit  in the use of the word  {\it finding}. It is  clear, that in the absence of such  measurement, the system will remain  in the state  given by the unitary evolution of the state $|B \r$, namely $e^{iHt}|B \r$. The probability interpretation is  only valid in  conjunction with a measurement.   An alternative posture, supporting the  notion  that one  can consider the system  described  by a state $|\psi\r$
 to have  a definite value
 (even if  it is unknown to us)  for quantities  for which $|\psi\r$
 is not an eigenstate,  is known to lead to conclusions that  are  in conflict  with  Bell«s  inequalities (See \cite{Mermin}),  and thus to contradiction  with experiments  (see also  the discussion in  Section IV A). 
 
 Getting back to  the analysis of Schwinger process,  we can use it to  further clarify  the situation we face in the cosmological context.  First, let us  note  that we must imagine the electric field to be turned on at some finite time, for otherwise, the problem of  the electron field in interaction with an external electric field, would be stationary, and if the system was prepared in its vacuum state, or  the state corresponding to the minimal eigenvalue of the full hamiltonian, it would remain in that state for all times and the issue of pair creation would not  make any sense.  So, let us assume that the electric field is  turned on  adiabatically during an  interval $\Delta (t)$ centered at  some $t_1$. Analogously, we assume the electric field to be turned off adiabatically  around some later time  $t_2$. Let us take  the electric field pointing in the  direction $x$.  We now assume that the system is prepared at  a time $t<<t_1$ in the vacuum state ( $|in,0\r$), and we ask about the probability amplitude  for observing the system in the state containing an electron and a positron, in  the single  particle states $\psi_1$ and
   $\psi_2$, respectively.  This question has  a very well defined answer in quantum theory, which is  simply 
   \be
   \l0; in | S |(\psi_1, -), (\psi_2,+);out \r .
    \ee
     However, let us note that we can not assume that the system, in the absence of a measurement,
   has  a well defined  probability of being in the state,
   \be|(\psi_1, -), (\psi_2,+);out \r,  
   \ee
   among other things, because the initial state is invariant under translations in the $y, z$ plane,  the hamiltonian preserves this  invariance, but the state,
   \be
    |(\psi_1, -), (\psi_2,+);out \r 
    \ee 
   will not, in general, share such invariance.  Thus,  while  we are perfectly justified in viewing the $ S$ matrix calculation to yield the  prediction for probability for the observation of pair creation  out of the vacuum, when contemplating the {\it measurement} of the number of pairs  at  a certain time, we are not justified  in regarding the state of a field as  being anything but  $U| B \r $ (where U is the unitary time evolution operator), in the absence of a measurement. Similarly,  in the early universe within the  inflationary context, we can find  no justification  to  view the state  of the universe, as anything but the state $U| 0 \r $, in the  absence of a measurement. 
   
   One other notion that is  sometimes  mentioned in oral discussions (we have never seen this  proposal being advanced in any  written article, or book),  is that somehow what  accounts for the transition from the  early  H\&I universe to the late asymmetric  one, is  similar to a ``spontaneous symmetry breaking". In this regard, we must note  that in contrast with  what occurs,  say,  in the breaking of  the  rotational symmetry in a ferromagnet  as the temperature  drops  below the critical value, which is a process intimately tied  to  environmental perturbations,  which  play an important  role in  tipping the system from  an unstable to one of the thermodynamical  stable states,  in  the cosmological situation at hand  we lack the  thermal environment  with  its perturbations,  but also  the phase transition from the one  characterized  by the symmetric   Bunch Davies  vacuum, to  one  where  there are only asymmetric  (but  degenerate)  thermodynamical  stable states.  If  what  one  wants to consider is  something like  the spontaneous symmetry  breaking in field theories, one should note that  there are no arguments indicating that, in the cosmological situation at hand the ground  state is  one lacking  the rotational and translational symmetries possessed  by the   Bunch Davies vacuum. In fact,  if one  were to advocate  such views  one would have to point to  the Nambu Goldstone  bosons or  the Higgs  mechanism   which are known to  be intimately tied  with such breakdown, and which are of course nowhere to be found in the present context.  We refer the reader, interested  on the  general subject to  the extensive  discussion in \cite{SSB}.
   
Another, quite recent and   very intriguing proposal that  must  be mentioned,  is the  approach  based on  Bohmian  quantum mechanics\cite{Bohm},   as adopted in \cite{Valentini}.  Here  the issue  one  must come to terms,  in the context of the problem at hand,  is whether  one  assumes that  what " gravitates"  is  the "pilot wave" or  the individual   "configuration trajectory" ( i.e.  whether  the  energy momentum tensor is  computed  using one or the other). The  problem for us is that  there seems to be no satisfactory  choice:  If we take the view that  it is the pilot wave what gravitates, the fact that the  wave function  corresponding to the vacuum is    H\&I, leads  us to conclude the universe is H\&I  throughout its complete history.  If on the other had,  one  takes the view that  it is  the specific  configuration space  trajectory what  gravitates,   then the  universe  was  never  H\&I and  one would have  a hard  time trying to argue that  the appropriate wave function to be used  in the inflationary context is the Bunch Davies vacuum. There might be some way to  escape  this  conundrum, but  it  is  very hard to envision.
   
  Finally  we  should mention the proposals  of solution of the problem,  based on the  
   {\it decoherent  (or consistent ) histories approach} to the measurement problem in quantum mechanics\cite{ConsitentHistories} (a slightly modified  version of  the formalism is proposed in \cite{Laura}). The general scheme is based  on  the   consideration, given  a quantum state  of the system  
   $ | \Phi  \rangle$   (or more  generally a density matrix $\hat \rho$) for the system at time $t_0$,   of  families of histories   characterized  by  a set  of  projection operators $\lbrace \hat P_{n} (t_n)\rbrace $,  each of  which   is  associated with the system possessing a value of certain physical property  in  a given  range   at a given time  \footnote{In  the  cosmological  setting, one  must use  a  subtler  relational time  approach\cite{HartleCosmology},  where   one of the dynamical variables is used   as an effective time  parameter.   The cosmological  scale factor  is a popular choice.}.  That is,  each of the projector operators  is  associated  with a  certain range  within the spectrum of  a given observable.    A given family  $F $ of such projectors, is  called  self consistent, if the resulting histories  do not interfere  among themselves,  a situation that allows   the assignment of  probabilities  to  each  individual  `` coarse grained history"  within the family.  
   
  In fact, the  
   probabilities  to  be  assigned  to  one such   a particular  coarse grained   history  within  a consistent family  are given  by:
   \be\label{Prob}
   P = Tr ( \hat P_{n}(t_n) U(t_n, t_{n-1} ) \hat  P_{n-1} U(t_{n-1}, t_{n-2} )  ...... \hat P_{2}U(t_{2}, t_{1} ) \hat P_{1}U(t_{1}, t_{0} ) \hat\rho  U(t_n, t_0 )^\dagger ), 
   \ee
   where  the $ U$«s  stand for the  standard unitary evolution operators  connecting two  times,  of   the  usual Schroedinger«s  quantum mechanics.  
   This approach  has  gained   some  followers in  cosmology community,  but it has  been  subjected to  some strong criticisms  in the foundational  physics community\cite{CHCritics}. 
   
 The   main  problem is that,  although the  scheme  works  fine once  one  has selected the decoherent family,  there exist, in principle,  an infinitude of  such  decoherent families,   which   are  however  mutually  inconsistent, (i.e.  there  are elements of $F$ and  $F'$   that do interfere, and  thus $\lbrace F \rbrace\cup \lbrace F' \rbrace $ is not   decohering). This  is  addressed  by the  ``single family rule"  which indicates  one  should never consider more than one family at a time.
  The  issue  becomes  how to single out a  particular family to be  that from which   the   particular history  that   one  considers  as the  actual one,   is to be   chosen (it  seems  very  reasonable  that the fact that one  assigns  probabilities  within a family   indicates that  the interpretation must be  that one of the   histories  in that  family is actualized  in our  world. Otherwise, one  must  wonder   what these probabilities refer to (i.e.  the probabilities  assigned according to \ref{Prob},  are probabilities of what? (see however \cite{Hartle-Languaje}). Recall that we  do not want to say that they are probabilities of  observing a certain value  of a physical  quantity  when  that quantity is  measured, or an  observation is made,  because we  do not  want to  bring concepts  like  measurement or observation into the discussion.
 
    This  is,  there  seems  to be, in principle,   no clear  way to single out a specific   family without relying on  an {\it a-priori}   given  set of  questions one is  asking-- those associated  with the  quantities    whose spectral  family one  choses   to construct the family - and this   can lead  to serious  interpretational  difficulties, in general.     
    
     In  a given  experimental setup, we might  guide ourselves, in practice,  by the  questions the experimental set up is ``asking" (in fact, this  has a close  analogy  with the 
     use Bohr rule in a given  experiment or series of experiments). However, in the  absence of such  guidance (i.e. without  {\it a priori} considering that the experimental set up corresponds to asking  certain yes /no  questions,  as  it seems to  be required  if one  does  acknowledge  the  possibility of  all  superposition  states of the  apparatus itself)   one  does not know  know   how to select the family. Note  that  one   is not asking  how to  select a  particular  history  within the family, but  how to select  a particular the family from within the collection of all possible decoherent families.
  
 In  fact, it is hard  to see, in describing the universe,  what  would  dictate the selection of the appropriate  projector operators,   and
  thus  of   the appropriate  family, (if we require a description  which   do not  makes use  of our own  existence  and  our own asking  of  certain questions, as part of the input). 
 
  Thus  in the cosmological context  the problem  can  be  seen, for instance in the following  example: Consider  the family of projector operators as is  done in   \cite{HartleCosmologyNew},  and   obtain their results,  but then note that  alternatively  we might consider the   following  family: 
   Now  construct the projector operator into  the  space of  homogeneous and isotropic  states $P_{HI}$.  This  is  simply the  projector into the  intersection of  the kernels of the generators of  translations and   rotations.  Let us  further    define   $P_{non} \equiv I-P_{HI}$  the   orthogonal projector. Take  the initial state  for the quantum fluctuations (usually called the vacuum)   
  $|\Phi_0\rangle$,  and  note that  it is homogeneous and  isotropic. 
    
    Now  take any set  of values  for  time  $\lbrace t_i \rbrace$  and   consider the family  associated  with that initial state  and   the  two  projector operators  $P_{HI}$ and $ P_{non} $ at  all those  times.    This    can easily be  seen to define a  family of consistent histories,  simply because the   dynamics  preserves the  symmetries  (homogeneity and  isotropy). 
     
       Thus,  one  might consider the  question, what is the probability that (at  given  time, characterized in the appropriate relational way),  the universe is  isotropic  and homogeneous.
   This   can  be  evaluated   using the  formula \ref{Prob} starting with the   vacuum  state.
   
   Any  history  containing the orthogonal projector  at  any time  $P_{non} $, will have   zero  probability,  but the history containing  only  the  operators  $P_{HI}$   will have probability  one. This   leads us to conclude that  at any time the universe is homogeneous  and isotropic.  It thus  can  have  no inhomogeneities or anisotropies at all. We would have then to face  not only that   problematic conclusion, but also  the  fact that  the  approach has  lead  us  to  two  conflicting  conclusions.  This later one,  and the one obtained in,  say,  \cite{HartleCosmologyNew}. We  would thus  need to  find  a way to  decide  how  to believe one of them  and not the other,  despite the fact that both   are obtained by the  very same  procedure\footnote{After  the first  version of this  article was written the author learned that the advocated posture is  that one  should believe both,  and use  the appropriate one  in connection with the  questions one is asking (see \cite{Hartle-Languaje}), although this posture is not shared by the authors of \cite{Laura}, \cite{Laura2}. I  admit  I  simply can not  come to terms   with such  view of  the  nature of  science  and  of  the nature  of knowledge in  general. This  views  seemed to be shared for instance  by the authors of \cite{CHCritics}. The motivation for the approach advocated there, is  the quest  
   for a  formulation that is  consistent with  both  quantum theory and relativity, and  one can only agree with such goals. However in my view the price  is  just too heavy. }.
    Apparently  one   would  have to  adopt   a rather  problematic  position   of the kind of   the posture {\bf  b) } in  section  III,  about the  nature of quantum theory, with the difficulties already  mentioned there.   
     It  seems that this is not  a satisfactory situation regarding  something that  ought to  serve as  a fundamental theory and, in particular,  to help us deal  with the quantum aspects of the early universe. The interested  reader is referred to the  literature, particularly to the  works referred  above  for  much more extensive discussions on the  matter.

  
   \section{Addressing the Problem}
   
   Before   presenting these ideas, we should warn the reader  that these  are, at this  stage, far from being completely developed.  
   One might  be tempted to  be frustrated that after reading  about the shortcomings of the traditional  explanation, one is presented  with a  relatively  vague  set of ideas,  so we warn the reader against the fallacious  comparisons   between I) an  account that  is  presented  as `` fully understood and established", and  based on the  existing accepted  laws  of  physics,  and II)  a  research project under construction, which  seeks to  make the first steps in understanding what is  belief to be a new and  not understood aspect of physics,  which is  argued  must play a role  in addressing the shortcomings of the generally  accepted  paradigm. The first  should be judged, as  we  have,  as  a ``finished  product,  ready  for consumption", while the former  should be judged  on the basis of its potential to  successfully  deal  with the issues on  general grounds. In fact, this is presented here just as  one  example  of a possible path out of the  problem,  and  certainly not  as anything  that can be regarded  as even close to finished.   We  will,
    mention some of the most  serious hurdles,  found so far,  that the program  must  deal with before it can be  considered  as part  of the  ``understood  areas  of physics".  Actually, one of the aims of the present manuscript is  to encourage  other colleagues to  consider alternative proposals having the potential to address  the exposed problem.  Having clarified what the reader can expect in this section,  we can proceed  with the exposition.
   
     We have,  so far, seen that the existing accounts  of a transmutation  of   the
     homogeneous and isotropic  situation to  which inflation takes the universe into the  in-homogeneous  and  anisotropic  situation  containing  the seeds  of cosmic  structure are flawed. Moreover, we  have argued  that  one  can not hope to find  an  appropriate  justification within the established paradigms of physics.    The   general unease in  accepting  these conclusions,  and the many proposal and  heroic efforts  in attempting to escape  from them, are  quite understandable  in view of the   fact   that the specific  form of the `` inflationary  predictions"  regarding the  origin of the primordial seeds of cosmic  structure   work so well.    However  having  convinced  ourselves of the   severe  shortcomings  we face, it seems clear that one  should seek  to add  new  elements  to such paradigm  in order to deal  with the problem.  In other words,  we  must incorporate  some, hopefully  minimalist,  extra  element into  physics   which would justify the success of the  `` inflationary prediction" of the   primordial perturbation spectrum  observed  in the CMB. 
       In this section, we  will sketch a path  leading to a  proposal  for a possible  source of such  modification, inspired  on  the ideas originally expressed  by Penrose  in \cite{Penrose1}, indicating that  Quantum Gravity  might  be  associated  with a modification of  Quantum Mechanics  that  would  resolve the measurement problem.   It should be  stressed that this  is nothing  but a rough  description of  
     a path that can be   envisioned  at this time,  and that a  explicit calculation is far beyond the scope of this  work,  among other reasons,  because  we still do not have a fully workable theory of  quantum gravity  which should   be the starting point.

 As we  have seen, the measurement problem is   very much an open problem  if  we insist (as we do here)  that  quantum theory  should be regarded as a theory of the  world as it is , and  not merely as a tool
 that     is of some use  in making certain predictions (see discussion in section III). In fact  the problem  becomes  much more   serious  once  we   want to go  beyond  the consideration of   simple  quantum mechanical examples and  consider  applying  the   usual interpretational  recipes  to   actual candidates for  fundamental  physical  degrees of  freedom  characterized, for instance,  by  quantum fields.  The attempts to do so  lead  to  very severe conflicts  with  standard  notions  of causality as  revealed  by  the  analysis of R. Sorkin\cite{Sorkin}.   There, it is  shown, that if  we
 take  the view that a projection into a given subspace of a Hilbert  space can  be  generically triggered  by   the decision to  carry out a  
 measurement,  the result is  that information can be transmitted faster than light.  As the author puts it,  we  end  up  with a  EPR type of situation  in  which the normal  impediments  for   a-causal messaging  are not  at work.
  This  analysis  clearly  leaves us  with  a  situation in which  we have no
   general interpretational framework  for  the  quantum  field  theories in general,  and   in particular, for any such theory  involving the gravitational degrees of freedom.

  In this context  we present   briefly vision  of  how the proposals  first made in\cite{Us}  might fit into the wider structure of our physical theories.
 The  basic  idea  we want to contemplate here  is to related the problem at hand   to   a different  problem  encountered  when trying to write  a theory of  quantum gravity through the canonical  quantization procedure.  It is  well known that,  when  following  such path either 
 in  the old  Wheeler  de Witt approach\cite{WdW},  or its  more  modern incarnation, in the form of Loop  Quantum Gravity\cite{LQG},  one  ends up  with  an atemporal theory.  This is  known as {\it the problem of time in quantum gravity} \cite{TIME}. That is,  in   both  schemes,  one  starts  with a formulation  in which the basic   canonical variables  describe the  geometry of a 3- spatial  hypersurface  $\Sigma$, and   characterize  the  embedding of
 this  3-surface in a 4-dimensional  space-time.   Let us  denote  generically  the  pair of canonical variables  as $({\cal G},  \Pi )$.
 However time, or its  general relativistic  counterpart,  a time function usually specified by the lapse  function and shift vector,  is no longer part of the theory. The  
most hopeful approach towards  addressing this  problem is to  consider,  simultaneously  with the  geometry, some  matter fields, which  are canonically  described  by a  set of  ordered pairs of variables
 \be \lbrace(\phi_1, P_1),...., (\phi_n, P_n)\rbrace,  \ee  and  to identify   an appropriate  variable $ T(\phi_1, {\cal G}, \Pi)$,  in the  joint  matter  gravity theory,  that  could  act  as  physical  clocks  and  to  characterize  the  state  for the remaining  variables, in terms of the correlations of  their values  with those of the physical clock. 
That is,  one  starts  from  the   wave  function  for the  configuration variables of the theory 
$  \Phi(\phi_1,..,\phi_n, {\cal G})$, which  must   satisfy the  so called hamiltonian  and  momentum constraints
 $ H_\mu   \Phi(\phi_1,..\phi_n, {\cal G} ) =0; \mu = 0, 1, 2, 3.$.  Next,  one  needs to  obtain  an effective  wave function $\Phi'$ for the remaining
   variables  by  projecting $\Phi$ into  the  subspace  where  the operator $T(\phi_1, {\cal G},\Pi)$  takes a certain range of values.  That is, let us denote  by $P_{T,  [t  ,  t+\delta t]} $  the projector operator  onto the subspace corresponding that part of spectrum of the operator $T$ lying   between the values   $t $ and  $ t+\delta t$.  The  next  steep  would be   to   recover a Schr\"odinger  type of   evolution  equation by   studying the dependence of  $ \Phi' (t)  \equiv P_{T,  [t  ,  t+\delta t]} \Phi' $ on the  parameter  $t$.   Needless is to say, that the  actual construction, in the  general  relativistic setting,  would need to overcome  the more   general character of the nature of the  time parameter, which would need to be specified  as  a global time function  ${\cal T} $ which  one  might  want  to  map  in the reconstructed  approximate space-time. That is,  after  obtaining, by the above  procedure, a wave function associated  with  the spectrum of the operator of ${\cal T}$,  we  might use  it  to  compute   the  expectation values of the   3 dimensional geometrical operators  (say the  triad  $E^a_i(x) $ and  connection $A_{a}^i (x)$ variables of Ashtekar, or  some  relevant   smoothing thereof) for  the wave function $ \Phi' (t)$. Such collection of  quantities  could be  seen as providing the   geometrical descriptions of the  ``average"  space-time in terms of the  $3+1$  decomposition of space-time.    In other words,  one  would   have  constructed  a  space-time  where the slicing  would  correspond to the   hypersurfaces   on which   the  geometrical    quantities   are given  by the expectations  of the projected  wave  functions $ \Phi' (t)$, and  thus  one  would  be able  to characterize the   space-time  and  its slicing in terms of   the   lapse and shift  functions. 
 The precise realization of this procedure  depends  strongly on the situation and  specific   theory  of  matter   fields  which  one  is  considering,  and  such  study is  quite  beyond  the scope  of the present paper,  and on the other hand  several works  along these  lines exist in the literature \cite{RelationalTime}. 
  The point, however, is that  the standard  Schr\"odinger type  equation  emerges only as an effective description, and is only approximately  valid.  Under those  circumstances, small  modifications  of that equation would  not  be  unexpected.  In fact, in a recent analysis \cite{Gambini-Pullin} of the  changes  in such  effective-Schr\"odinger-time evolution  equation  for a  quantum mechanical  system,  when described in terms of the time   measured  by  a physical clock, it was found  that  the  modifications  do  away  with the  unitary evolution. We  believe, therefore,  that  this can be the grounds where a modification of  Schr\"odinger evolution, involving  something akin to a collapse of the wave function, might find  its explanation.  There  are,  indeed,  several proposals  for such a modification centering on the analysis  of  standard  laboratory  situations \cite{Diosi, GRW, Otro}.  
  It  seems that  with   a paradigm where the  quantum  jumps occur   generally,   rather than  being triggered by the decisions of observers to measure such and such  quantity,  we  might  avoid  the sort of problem discussed in \cite{Sorkin}.
  
      Going  back to our situation, it is clear that  when  considering, in conjunction with  the quantum  evolution of the  matter fields,   the semiclassical  approximation in the  gravity  sector, it is clear one  would need  to  incorporate,  in some form, the   back reaction of the  metric  variables  to the  quantum jumps of the  wave function of  matter.  Needless is to say,  that nothing of this  sort  can be  achieved  at present time,  among other reasons  due  to the fact that we  lack a  truly workable theory of quantum gravity. 
      Nevertheless  one can envision a setting  leading up to the  kind  modifications  to the standard physics that we think are  needed if we want  to  successfully address  the issues  we have  been discussing in the  previous sections.
 Focussing more concretely  on the   problem  at  hand,  namely the  quantum origin of the seeds  of cosmic  structure  during inflation,
   one  can make  an  educated  guess about    some  of the characteristics  of an   effective  description  emerging  from a procedure  like  the one envisaged  above,  and starting with  a appropriate theory of quantum gravity. With the  view on its  application to addressing  the   problems  we have  encountered  in considering the   quantum origin of the seeds  of cosmic   structure   during inflation,  a proposal in this  spirit   has been put forward  in\cite {Us}.

In accordance with the ideas above, and  further motivated  by Penrose's  proposals   that quantum gravity should  play a role  in triggering a physical collapse of the wave function  for matter fields, the scheme  was built to rely on  a semi-classical description of gravitation in interaction with 
quantum  fields as reflected  in the semi-classical Einstein's 
equation 
\be\label{EE}
 R_{\mu\nu} -(1/2) g_{\mu\nu} R =8\pi G  \l T_{\mu\nu}  \r 
 \ee
  whereas the
matter  fields are treated in the standard quantum field  theory in curved space-time fashion. 
It  is clear that this can not be  but an effective  description of limited  applicability,  but we  will assume that  it includes the cosmological  setting at hand. The extra element that  we consider, as reflecting the ideas  discussed in this  section, is  the quantum  collapse of the   wave function  of the 
matter fields.  That is,  the   normal unitary evolution, characterizing  the field  theory, will be supplemented  by  instances of  quantum collapse   of the wave function, thought to   be triggered,  somehow,  by the  effects  of gravitational degrees  of freedom  that are not described  by the metric  (the  metric  is  regarded  as  a mere effective description  of some  average  behavior of the true  gravitational degrees of freedom). The  rest  of  the system  is  then  described in terms of the Heisenberg  picture, where the field  operators  reflect the standard  and   unitary   part of the evolution,  while the  states will remain constant,  except  when  a collapse occurs,  which will be characterized by  a  random jump  of the state  $ |\Phi ' \rangle $  to  one  among a  set of suitable   related  states
  \be \lbrace |\Xi'_1 \rangle, .....,|\Xi'_i \rangle,.... \rbrace,
   \ee 
  in  what  we  will call   a ``self induced  collapse of the  wave  function" :
\be\label{Col}
|\Phi ' \rangle  \to |\Xi'_i \rangle 
\ee
 Alternatively, we might  think of  the formalism as describing a sort of {\it Interaction Picture}, whereby  all the  standard  part of the evolution is  encoded in the field  operators,  while  the  collapse  is considered as  the  effective description of the  nontrivial interaction with the fundamental  quantum gravity degrees of freedom.  It is  clear that   the  equation  \ref{EE}  would not  hold, in general,  through  the collapse of the  wave function.  At such  times,  the excitation of the  fundamental quantum gravitational degrees of freedom should be taken into account, with the corresponding breakdown of the semi-classical approximation. The  possible breakdown  of the  semi-classical approximation is  formally represented  by the inclusion of  a  term  $Q_{\mu\nu} $ in  
 the semi-classical Einstein's   equation, which is supposed to become nonzero {\bf only}  during the  collapse
  of the quantum mechanical wave function of the matter fields. Thus, we write,
\be
\label{SemiCEQ}
  R_{\mu\nu} -(1/2) g_{\mu\nu} R +Q_{\mu\nu}  =8\pi G  \l T_{\mu\nu}  \r .
 \ee
 Thus, we consider the development of the state of the universe during the time at which the seeds of structure 
 emerge to be initially  described by a H.\& I. state for the gravitational and matter D.O.F..  At some point, the quantum state of the matter fields reaches
  a stage whereby the corresponding state for the gravitational D.O.F. leads to  a quantum collapse 
  of the matter field wave function.
   The resulting state of the matter fields  does no  longer  need to share  the symmetries of the initial state, and
  its connection to the  gravitational D.O.F.,  which is  assumed to again  be accurately described by Einstein's semi-classical  equation,  leads to  a  geometry that is no longer homogeneous and isotropic.

 Concretely, the
starting point  of the analysis,is,  as usual,  the action of a scalar field coupled to
gravity,
\begin{equation}
\label{eq_action}
S=\int d^4x \sqrt{-g} \lbrack \frac {1} {16\pi G} R[g] - 1/2\nabla_a\phi
\nabla_b\phi g^{ab} - V(\phi)\rbrack,
\end{equation}
 where $\phi$ stands for the inflaton or scalar field responsible for inflation and $V$ for the 
inflaton's potential.

 It is customary to split both, metric and
scalar field, into a spatially homogeneous (`background') part and an
inhomogeneous part (`fluctuation'), i.e.   we write the metric  as $g=g_0+\delta g$,
while the   scalar field, which as we said  is to be treated  quantum mechanically  in a semiclassical approximation,   is  supposed to be  in a state $|\Phi_0\rangle$ (we  are  dropping  now the  primes that were  used to indicate effective nature of the  state as  discussed  above, but  of  course  the discussion remains in that  same  setting).  This  state   is   assumed to  be  homogeneous and isotropic  as  a result of  the  early  stage of inflation. 
 
 \smallskip
Next, we   define $\phi_0  (\eta) = \langle\Phi_0  | \hat\phi (\eta , x) |\Phi_0\rangle $  which is of course independent of $x$.

 \smallskip
 The  semiclassical   equations,  together with the field  scalar  field  equations (and  after suitable  renormalization of the  energy momentum  tensor),  then   lead to  a ``background  solution" that corresponds to the standard inflationary cosmology  which,   when written using the conformal time, corresponds to the   scale factor:
$
a(\eta)=-\frac{1}{H_{\rm I} \eta},
$
with $ H_I ^2\approx  (8\pi/3) G V$and with the scalar $\phi_0$ field in the slow roll regime so $\dot\phi_0= - ( a^3/3 \dot a)V'$ (  with  the `` dot"  representing  $\frac{\partial}{ \partial \eta} $   and 
  the  quantities  associated  with the scalar filed  being understood  as  expectation values evaluated in  the  state $ |\Phi_0\rangle $,  of course).

 The    emergence of  the  inhomogeneities and anisotropies  is then   addressed as  follows:  We   define the   shifted  quantum field $ \hat \delta\phi ( \eta, x) =\hat \phi( \eta, x) - \phi_0 ( \eta) \hat I $  where $\hat I $  stands  for the  identity operator in the  scalar field  Hilbert space.  The   state $|\Phi_0\rangle$,  when characterized  in terms of the field  $ \hat \delta\phi $,   is supposed not  only to  have  zero expectation value,  but  to  be essentially  the Bunch-Davies vacuum  associated with the  almost de-Sitter expansion (or  something  very close to it). Thus  
 \be \langle\Phi_0  | \hat \delta \phi (\eta , x) |\Phi_0\rangle = \langle\Phi_0  | \hat \delta {\dot \phi (\eta , x)}|\Phi_0\rangle  =0. 
 \label{VEV}
 \ee
The perturbed metric can be written as
\begin{equation}
ds^2=a(\eta)^2\left[-(1+ 2 \Psi) d\eta^2 + (1- 2
\Psi)\delta_{ij} dx^idx^j\right],
\end{equation}
 where $\Psi$  stands for the relevant perturbation and is called
the Newtonian potential.

The perturbative treatment   of the  above situation  using  the semiclassical Einstein's equations leads to  an  equation  connecting  the  metric  perturbation to the expectation value  of the  perturbed energy momentum tensor, which   lowest order take the form :
\begin{equation}
\nabla^2 \Psi  = 4\pi G  \dot \phi_0  \langle\Phi |\hat \delta{\dot \phi}  |\Phi\rangle.
\label{main2}
\end{equation}

It is clear  from  \ref{VEV} that  for the original  state $|\Phi_0\rangle$,  we obtain $\Psi =0$, and thus the  metric  would be homogeneous and isotropic.  However, after the  state of the  field
  undergoes a series  of collapses  according to \ref{Col},  the  situation is   modified,   and, in  particular   the R.H.S. in   \ref{main2}   would become  different from $0$ and   generically  dependent on $x$,  leading to a  nontrivial  metric perturbation  characterized  here by the newtonian potential $\Psi\not =0$.  Thus  we  have the actual   emergence of the seeds of  structure   as a direct  result of the  dynamical quantum  collapse   of the wave  function. 
 The  details  of  various  aspects of this  analysis can be found  in \cite{Us}, and in \cite{othersus, Napflio, Adolfo},   where  it  is shown  that  the above  scheme, supplemented  by   a some ``natural assumptions"  about the  state after the  collapse,   leads  to a  prediction of the  power spectrum of the CMB fluctuations,  which matches  the observations, if the collapse  is assumed   to  occur in a particular  form\footnote{The standard prediction  is obtained when  the conformal time  for the  collapse of a mode  is  proportional to  its co-moving  wavelength.}.  We point the interested readers  to the  above mentioned  works  for a more detailed exposition of this  particular realization of the general set of  ideas discussed in this section. Those  works  do constitute  a concrete  proposal (by no means  claimed to be the only possible one) which  explicitly addresses the problem  at hand, and  which, as a bonus,  leads to  concrete  predictions  that can, in principle,  be confronted with observations.  
 
   The  main obstacles  facing this  proposal, as  far as  can be  envisaged  at this point , are: 1) The collapse  schemes  that is considered  here  in the cosmological context  must  be made compatible  with   a generic  solution to the measurement problem and  thus framed  in  a quantum field  theoretical implementation of the general type  considered in \cite{GRW, Diosi}.  2)The  complete  formal  development requires, as indicated  before, a fully  workable and  satisfactory  theory of quantum gravity.
        Some particular  issues that  will doubtless  become relevant are  related  with the compatibility  of schemes involving a the  collapse of  the  wave  function, with general covariance.
         Here we should  note  that the tension is  already there in  discussions of  Einstein-Podolsky-Rosen-Bohm Gedankenexperiment, (EPR) type situations. In fact, although the  standard  EPR set ups  are known no to allow  faster  than light communication, the EPR-gone bad  schemes discussed  in \cite{Sorkin} show  that  there are  aspects of these problem  that  are still not fully   understood.   It is  thus  clear  that  the path required for a full development of the program is  long and difficult. On the other hand  the possible   connection  between of all these  issues, as envisaged  by this program   is  evidently  filled with  enormous potential.

         \section{Conclusions}
  
  In the cosmological setting,
we seek an historical account. That is, a  development in time   that describes the  cosmic evolution, and which follows the laws of physics.
Such description  should explain how did WE arise, in a path covering the emergence of the primordial density fluctuations, of galaxies, stars, and  planets,  and eventually living organisms, humans, cultures, etc..
Such an account should not rely on the measurement  (in) abilities of the late evolved creatures to explain the  emergence of  conditions that make them possible.  From this perspective,  one can not justify identifying some D.O.F as irrelevant environment, based on the current, or even permanent,  limitations of humans, in  the analysis of the emergence of the primordial  density fluctuations, for doing so  leads to a circular argument with no explanatory value.  Alternative postures such as  that in 5b)  would make the use of quantum mechanics unjustified, in general, as discussed there.
  
    One can not take one position regarding quantum mechanics (or any theory) in one instance and a different one in another.
    
     It is still a remarkable fact that the HZ spectrum coincides with the calculations of the uncertainties in the evolved quantum state of  the inflaton  field.  But we need to understand why is that.
     
  There are, of course, alternatives, but those require the introduction of some novel aspect in physics,  such as a ``dynamical collapse of the wave function",  along the proposals by Penrose \cite{Penrose1},  Diosi\cite{Diosi},  GRW\cite{GRW}, and \cite{Otro}. We have shown  one simple implementation of these ideas in the exploratory work \cite{Us}, with further analysis in \cite{Adolfo}.
  
 One might wonder what is  gained  by  introducing  an un-explained, and  rather  ad-hoc  element into physical theory    versus  the  acceptance of  an  problematic   step  in the argument within   the established theoretical paradigm? Answer: The first thing is the explicit  acknowledgment that there is something missing, and that it  hould be  the subject of  further research. Second, the  modeling  and parametrization of the  unknown physics can be taken as the  starting point of the  research into  that aspect of physics  not yet incorporated into our fundamental theories. Finally,  the fact that cosmology offers  us  a large amount of  relevant data indicates that we are in  the position to  engage in a serious  scientific  study  about some of the characteristics of that new physics. 
   
    Therefore,  one  must conclude that, in facing this issue, physicists face a dilemma:  Take  
   as  a sort of  confirmation of  the  predominant   ideas  on the subject,  the agreement of the  calculations with the  observations,  and ignore the fact   that  the scheme can not be fully justified within the context of the interpretations provided by our current physical theories, or  admit the  existence of  shortcomings  and start to work,  towards  clarifying the issues,  and in the process hopefully gain some new insights in our quest to understand the functioning of our world. 
There is no doubt that in 
taking the second  approach we will be often  led to consider  erroneous proposals, and  explore paths  that turn out to be dead ends, (as could  clearly be the case with our own works mentioned  above) but rather than see this as a deterrent,  I  believe we  should take heed from Sir  Francis Bacon's profound observation\footnote{As cited  in Thomas Kuhn's ``The structure of Scientific Revolutions"}  about scientific methodology: {\it ``truth emerges more readily from error that from confusion"}. It seems  that if we  want to  stay true to the  scientific  spirit  the option is simple.

\section*{Acknowledgments}

\noindent
 This work was supported in part  by DGAPA-UNAM
IN119808-3. The author acknowledges very useful discussions  with (in chronological order) A. Perez,  A.  De Un\'anue,  Elias Okon , A. Bassi, B.  Kay,  M. Castagnino,  R. Laura  and
J. Hartle.  He, also thanks  R. M. Wald  for pointing out very useful and  relevant references.


\begin{thebibliography}{99}

\bibitem{Polarski-Kiefer}  ``Why do cosmological perturbations look classical to us? ", C. Kiefer \& D. Polarski, e-Print: 0810.0087 [astro-ph].

\bibitem{Cosmologists}
``Decoherence in Quantum Cosmology", 
J.J. Halliwell,
{\it Phys. Rev. D}, {\bf 39}, 2912,(1989);
``Origin of Classical Structure From Inflation",
 C. Kiefer
{\it Nucl.\ Phys.\ Proc.\ Suppl.\ } {\bf 88}, 255 (2000)
[e-Print:astro-ph/0006252];
 ``Semi-classicality and decoherence of
  Cosmological perturbations",
   D. Polarski and A.A.
  Starobinsky,{\it Class.\ Quant.\ Grav.\ }  {\bf 13}, 377 (1996)
  [e-Print: gr-qc/9504030];
`` Environment Induced Super selection In Cosmology",
 W.H. Zurek, Environment Induced Super-selection In Cosmology in
  {\it Moscow 1990, Proceedings, Quantum gravity} (QC178:S4:1990), p. 456-472.
 (see High Energy Physics Index 30 (1992) No. 624);
 ``Gauge Invariant Cosmological Perturbations" 
 R. Branderberger H. Feldman and V. Mukhanov,{\it  Phys. Rep.} { \bf 215}, 203, (1992);
``Decoherence Functional and Inhomogeneities in the Early Universe",
 R.  Laflamme and A. Matacz {\it Int.\ J.\ Mod.\ Phys.\ D } {\bf 2}, 171 (1993)
[e-Print:gr-qc/9303036];
 ``The self-induced approach to decoherence in cosmology,''
  M. Castagnino and O. Lombardi, {\it  Int. J. Theor. Phys.} {\bf 42}, 1281, (2003),
  [e-Print:quant-ph/0211163];
  ``Decoherence during inflation: The generation of classical inhomogeneities,''
  F.~C.~Lombardo and D.~Lopez Nacir,
 Phys.\ Rev.\ D {\bf 72}, 063506, (2005)
  [e-Print:gr-qc/0506051];
  ``Inflationary Cosmological Perturbations of Quantum Mechanical Origin"
J. Martin, {\it Lect.\ Notes Phys.\ } {\bf 669}, 199 (2005)
  [e-Print:hep-th/0406011];
``Best Unbiased Estimates for Microwave background Anisotropies", 
L.P. Grishchuk and J. Martin,
 {\it Phys.\ Rev.\ D} {\bf 56}, 1924, (1997)
  [e-Print:gr-qc/9702018];
  ``Decoherence in Quantum Cosmology at the onset of Inflation",
  A. O. Barvinsky, A.Y. Kamenshchik, C. Kiefer, and
I.V. Mishakov, {\it Nucl.\ Phys.\ B} {\bf 551}, 374, (1999)
  [e-Print:gr-qc/9812043];
  ``Quantum To Classical Transition
  of Cosmological Perturbations for Non Vacuum Initial States", 
  J.  Lesgourgues, D. Polarski and A. A. Starobinsky
  [e-Print: gr-qc/961101904030] .
 
 
 \bibitem{Padmanabhan} Section 10.4, page  364 of  `` Structure Formation in the Universe", T.
  Padmanabhan (Cambridge University Press, 1993).
  
 \bibitem{Weinberg} Page 476 ``Cosmology", S.  Weinberg (Oxford University Press, 2008).

 \bibitem{Mukhanov} Page 348 of ``Physical Foundations of Cosmology",  V. Mukhanov (Cambridge University Press, 2005).
 
\bibitem{Quantum Cosmology} ``Quantum Cosmology  Problems for the 21${}^{st}$ Century", J. B.Hartle, [e-Print:
gr-qc/9701022];
``Generalized Quantum mechanics for Quantum Gravity", J. B. Hartle,
[e-Print: gr-qc/0510126].

\bibitem{Inflationary Cosmology}  ``Quantum Mechanics of the scalar field in the new
  inflationary Universe",
  A. Guth and S.-Y. Pi {\it{ Phys.  Rev.  D}} {\bf 32}, 1899, (1985);
 ``Fluctuations in the Inflationary Universe",
S. W. Hawking {\it{Nucl. Phys.}} {\bf B 224}, 180,  (1983);
``Origin of Structure in the Universe"
J.J. Halliwell and S. W. Hawking,
{\it Phys. Rev. D}, {\bf 31}, 1777,(1985).


\bibitem {Us}
``On the Quantum Origin of the Seeds of Cosmic Structure"
 A. Perez, H. Sahlmann, and D. Sudarsky,   {\it Classical and Quantum Gravity}  {\bf 23}, 2317, (2006) e-Print: gr-qc/0508100.


\bibitem{othersus} ``The Seeds Of Cosmic Structure as a Door To New Physics",
proceedings of the conference "Recent Developments in Gravity NEB XII ", Napflio Greece,  June 2006.
 {\it  J. Phys. Conf. Ser.} {\bf 68}, 012029, (2007). 
e-Print: gr-qc/0612005;
``A path towards quantum gravity phenomenology",
proceedings of the conference   "ERE 2006", Palmas de Mallorca, Spain,  Sep  2006.
 {\it J. Phys. Conf. Ser.} {\bf 66}, 012037, (2007);
``A signature of Quantum Gravity at the Source of the Seeds Of Cosmic Structure",  
proceedings of the conference ``DICE 2006", Piombino, Italy,  Sep 2006.
{\it J. Phys. Conf. Ser.} {\bf 67}, 012054, (2007) 
e-Print: gr-qc/0701071; 
``The Seeds Of Cosmic Structure as a door to  Quantum Gravity Phenomena",  proceedings of the
 conference `From Quantum to emergent Gravity:Theory and  Phenomenology",  SISSA, Trieste,  Italy,  June  2007. e-Print: gr-qc/0712.2795. 

\bibitem{Adolfo} ``The seeds of cosmic structure: A phenomenological approach" A. De Unanue \& D. Sudarsky,  \emph{Phys. Rev. D} {\bf 78}, 043510 (2008)
 e-Print: gr-qc/0801.4702.

\bibitem{Guth}
 ``Quantum Mechanics of the scalar field in the new
 inflationary Universe", 
  A. Guth and S.-Y. Pi {\it{ Phys.  Rev.  D}} {\bf 32}, 1899, (1985);
   ``Fluctuations in the Inflationary Universe",
S. W. Hawking {\it{Nucl. Phys.}} {\bf B 224}, 180,  (1983);
``Origin of Structure in the Universe" 
J.J. Halliwell and S. W. Hawking,
{\it Phys. Rev. D}, {\bf 31}, 1777, (1985).

\bibitem{CMB} 
``Cosmological parameters From First results  of Boomerang"
A. E. Lange {\it et. al.}
{\it  Phys. Rev. D}, {\bf  63}, 042001, (2001);
G. Hinshaw  {\it et. al.}, {\it Astrophys. J. Supp.}, {\bf 148}, 135,
(2003);
``Power Spectrum of Primordial Inhomogeneity Determined
 from four Year COBE DMR SKY Maps",
 K. M. Gorski {\it et. al.}
  {\it Astrophys. J.} {\bf 464}, L11, (1996);
 ``First Year Wilkinson Microwave Anisotropy Probe (WMAP) Observations: Preliminary Results"
C. L. Bennett {\it et. al.} {\it   Astrophys. J. Suppl.}  {\bf 148}, 1, (2003);
``First Year Wilkinson Microwave Anisotropy Probe (WMAP) Observations: Foreground Emission",
 C. Bennett {\it et. al.}  {\it Astrophys. J. Suppl.} {\bf 148}, 97, (2003).

\bibitem{MPQM}  For reviews abut the various approaches to 
the measurement problem in quantum mechanics see for instance
 the classical reference ``Philosophy of quantum mechanics. 
 The interpretations of quantum mechanics in historical perspective" M. Jammer, 	
(John Wiley and Sons, New York  1974); 
 ``The Interpretation of Quantum Mechanics" R. Omnes, ( Princeton University
Press 1994), and the more specific critiques
`` Why Decoherence has not Solved the Measurement Problem: A Response to P. W. Anderson" -
S. L.  Adler {\it Stud. Hist. Philos. Mod. Phys. }{\bf 34}135-142 (2003), [e-Print: quant-ph/0112095];
``Why modal interpretations of quantum mechanics don't solve the measurement problem".
A. Elby, {\it Found. of Phys. Lett.}
{\bf  6},  5-19 (1993)
 and  the review of approaches to the problem presented
  in  A. Bassi  \& G. C. Ghirardi, {\it  Phys. Rept },{\bf  379}, 257, (2003)[e-Print: quant-ph/0302164];


\bibitem{QM as applied only to ensembles}
 ``The statistical interpretation of quantum
 mechanics" L.E. Ballentine {\it  Rev. Mod. Phys} {\bf .42}, 358-381,(1970).

\bibitem{Single System} ``Measurement of the Schr\"odinger wave of a single particle",
Y. Aharonov, L. Vaidman 
{\it Phys. Lett. A }{\bf 178} 38,(1993); 
e-Print: hep-th/9304147

\bibitem{Asher Peres} Sec 2.7 of  ``Quantum Theory  Concepts and Methods", A. Peres (Kluwer Academic Publishers 1993)

\bibitem{Zeh} ``Roots and fruits of Decoherence ", H. D. Zeh, in {\it Proceedings of the   Seminarire Poincar\'e}, Nov 2005 (eds. T. Damour B. Duplantier \&  V Revasseau,  Birkh\"auser, 2006); quant-ph/0512078. 

\bibitem{Hartle-objective} ``Quantum Mechanics of Individual Systems",  J. B. Hartle {\it Am. J. of Phys.} {\bf 36}, 704, (1968).

\bibitem{Measurement}  ``The problem  of Measurement",  E. Wigner {\it Am. J. of Physics}
{\bf 31}, 6, (1963);  Macroscopic quantum Systems and the quantum theory of measurement",
 A. Laggett {\it Prog. Theor. Phys. Suppl.} {\bf 69}, 80, (1980).
 
 \bibitem{Anandan} ``Quantum measurement Problem and the Possible Role of the gravitational Field"
 J. Anandan,  gr-qc/9808033.
 
\bibitem{Penrose1}  
``The Emperor's New Mind",
 R. Penrose, {\it The Emperor's New Mind}, (Oxford
  University Press 1989); R. Penrose, On Gravity's  Role in Quantum State Reduction,
  in {\it  Physics meets philosophy at the Planck scale} Callender, C. (ed.) (2001).

\bibitem{Problems Big Bang}  ``Inflationary universe: A possible solution to the horizon and flatness problems"A. Guth
 {\it Phys. Rev. D}{\bf  23}, 347, (1981). For a  more exhaustive  discussion see for instance the relevant chapter in
  ``The Early Universe", E.W. Kolb and M.S.  Turner, Frontiers in Physics Lecture Note Series (Addison Wesley
  Publishing Company 1990).

\bibitem{FAAP} ``Against Measurement",
J. S. Bell, {\it Phys. World} {\bf  3}, 33,(1990).

\bibitem{Hartle1} ``Quantum Cosmology  Problems for the 21${}^{st}$ Century", J.~B.~Hartle, [e-Print:
gr-qc/9701022];``Generalized Quantum mechanics for Quantum Gravity", J. B. Hartle, 
[e-Print: gr-qc/0510126].

\bibitem{Nuevos  QM}  For a  recent discussion  on the status  of  the subject see for instance 
``The Interpretation of  Quantum mechanics:   where  do we stand?" G. C. Ghirardi,  [e-Print quant-ph/ 0904.0958]

\bibitem{HZ} ``Fluctuations at the threshold of classical cosmology" E. R. Harrison, {\it Phys. Rev. D}, {\bf 1}, 2726, (1970); Y. B. Zel\'{}dovich {\it  Mon. Not. Roy. Astron. Soc.} {\bf 160}, 1p (1972). 

\bibitem{GRW} 
``A Unified Dynamics For Micro And Macro Systems",
G. C. Ghirardi, A.  Rimini, and T.  Weber,{\it  Phys. Rev.} {\bf D 34}, 470, (1986);
 ``Dynamical Reduction Models: present status and
  future developments", A. Bassi (2007) {\it Preprint} quant-ph/0701014v2.;
  ``Dynamical Reduction Models", A. Bassi and  G .C. Ghirardi  (2003) 
  {\it Preprint} quant-ph/0302164v2.


\bibitem{Adler} 
``Why Decoherence has not Solved the Measurement Problem: A Response to P.W.
Anderson", S. L. Adler, { \it Stud. Hist. Philos. Mod. Phys.}{\bf 34},135, (2003).

\bibitem{Neumaier} A. Neumaier, in the ``theoretical physics  FAQ" at his
University of Wien, webpage:  http://www.mat.univie.ac.at/~neum/physics-faq.txt. 
      
\bibitem{Schlosshauer} 
 ``Decoherence, the measurement problem and interpretations of quantum Mechanics" 
M. Schlosshauer, { \it Rev. Mod. Phys. }{\bf 76}, 1267 (2004), [e-Print:  quant-ph/0312059, page 9].
   
 \bibitem{Joos}   E. Joos, in "Elements of Environmental Decoherence", proceedings of the conference 
 Decoherence Theoretical, Experimental and Conceptual Problems" (Eds. P. Blanchard, D. Giulini, E. Joos, C. Kiefer and I. O. Stamatescu, Springer 1999) [e-Print: quant-ph/9908008].
  
\bibitem{Bell} 
``Speakable and Unspeakable in Quantum Mechanics", 
J. S. Bell (Cambridge University Press 1987).

\bibitem{Aspect} 
``Experimental realization of Einstein-Podolsky-Rosen-Bohm Gedankenexperiment:
A New violation of Bell's inequalities"
 A. Aspect, P. Grangier, G. Roger,
{\it Phys. Rev. Lett.} {\bf 49}, 91, (1982);``A New violation of Bell's inequalities"
 A. Aspect, P. Grangier, G. Roger,
{\it Phys. Rev. Lett.} {\bf 49}, 9, (1982).


\bibitem{RoadToReality} ``The Road to Reality",  R. Penrose (Alfred  A. Kopf- New York, 2006).

\bibitem{DEspagnat}``A note on measurement", B. D'Espagnat, {\it Phys. Lett. A} {\bf 282}, 133, (2001). 



\bibitem{EPR}  See for instance discussions about  the  EPR experiment in  A. Peres `` Quantum Theory: Concepts and Methods" (Kluwer Academic  Publishers, 1993)

\bibitem{Mermin}
 ``Is the Moon There when nobody Looks?"
 D. Mermin {{\it Physics  Today }} {\bf 32}, 38, (1985).
 
\bibitem{QSuicide}``The Interpretation of quantum mechanics: Many worlds or many words? "
M. Tegmark,  
(Fundamental Problems in Quantum Theory, Eds. M.H. Rubin and Y.H. Shih.) 
{\it Fortsch. Phys.} {\bf 46}, 855, (1998) 
[e-Print:quant-ph/9709032].

\bibitem{MWI} For  further discussion of the issue and  other points of view  see  for instance:
``Against Many Worlds Interpretations",
A. Kent,  {\it  Int. J. Mod. Phys.} { \bf A5}, 1745,(1990). 
[e-Print: gr-qc/9703089]; ``Against `Against many worlds interpretations'."
T. Sakaguchi, [e-Print: gr-qc/9704039]; ``Many worlds in one"
J. Garriga, A. Vilenkin, {\it  Phys. Rev.} {\bf D64}, 043511, (2001) 
[e-Print:   gr-qc/0102010];
``How many new worlds are inside a black hole?"
C. Barrabes, V. P. Frolov, 
{\it Phys. Rev.} {\bf D53}, 3215, (1996)
[e-Print:  hep-th/9511136].

\bibitem{B. Kay} In  the write up of this subsection  the author benefited from an extensive discussion with B. Kay.

\bibitem{Napflio}
 ``The Seeds of Cosmic structure as a door to New Physics" D. Sudarsky, 
{\it J. Phys. Conf. Ser.} {\bf 68}, 012029, (2007)
[e-Print: gr-qc/0612005] .

\bibitem{HollandsWald}  ``Quantum field theory is not merely quantum mechanics applied to low energy effective degrees of freedom",
S. Hollands \& R. M. Wald,
{\it Gen. Rel. Grav.} {\bfÊ36 }, 2595, (2004). 
[e-Printe:  gr-qc/0405082].

\bibitem{Starobinski}``Semiclassicality and decoherence of
  Cosmological perturbations", D. Polarski and A. A.
  Starobinsky, {\it Class. Quant. Grav. }, {\bf 13}, 377 (1996)
  [e-Printe: gr-qc/9504030].


\bibitem{Squeezed} ``Sub-Poissonian processes in quantum optics"
 L. Davidovich, {\it Review of Modern Physics} {\bf 68},  127, (1996);
``Squeezed states of light", D. F. Walls
{\it Nature} {\bf 306}, 141, (1983);
 See also pg. 4 of  ``Lectures to Quantum Optics", W. Vogel \& G.G. Welsch, (Akademie Verlag,
Berlin/ VHP Publishers, New York, 1994)

\bibitem{correlations -squeezed} ``Spin Squeezing and Light Entanglement in Coherent Population Trapping" A. Dantan, J. Cviklinski, E. Giacobino, and M. Pinard
 {\it Phys. Rev. Lett.} {\bf 97}, 023605, (2006).


\bibitem{Mott} N. F. Mott  ``The  Wave  Mechanics of  $ \alpha$- Ray tracks", {\it Proc. of the Royal Soc. of London} {\bf 126} No 800, pg 79, (1929).

\bibitem{Jerome 1} 
``Inflationary perturbations: The Cosmological Schwinger effect"
J. Martin, [e-Print: hep-th/07043540].

\bibitem{SSB}``On the symmetry of the vacuum in theories with spontaneous symmetry breaking", A. Perez  \& D. Sudarsky. e-Print: e-Print:0811.3181 [hep-th].

\bibitem{Sorkin} ``Impossible measurements on quantum fields'', R. D. Sorkin, in ``Directions in General Relativity" (B. L. Hu,
 T. Jacobson Eds. Cambridge Univ.\ Press,
 Cambridge, 1993).

\bibitem{WdW} ``Quantum Theory Of Gravity. 1. The Canonical Theory'',  B. S. Dewitt,
 {\it Phys. Rev.} {\bf 160}, 1113 (1967);
``Superspace and the Nature of Quantum
Geometrodynamics'', 
 Battelle Rencontres: 1967 Lectures in Mathematics and Physics, (J. A. Wheeler,  in C. DeWitt and J.A. Wheeler, Eds.)
 W.A. Benjamin, New York, 1968.

\bibitem{LQG} See for example, ``Loop quantum gravity'', C.
 Rovelli, 
 {\it Living Rev. Rel.} {\bf 1}, 1 (1998)
 [e-Print:gr-qc/9710008];
 ``Quantum geometry and gravity: Recent advances'', A. Ashtekar,
[ e-Print: gr-qc/0112038];
 ``Introduction to modern canonical quantum general relativity'', T. Thiemann,
 [e-Print: gr-qc/0110034].


\bibitem{TIME}  B. S. De Witt {\it  Phys. Rev.} {\bf 160}, 1113,(1967); J.A. Wheeler in
{\it Battelle Reencontres 1987}  eds.  C. De Witt \& J.A Wheeler (Benjamin, New York,1968); ``Canonical Quantum Gravity and the problem of Time",  C.J. Isham {\bf GIFT Semminar}- 0157228 (1992) qr-qc/9210011

\bibitem{RelationalTime}  See  for instance, D.  N. Page \& W.K. Wootters {\it  Phys. Rev.} {\bf D 27}, 2885, (1983); W.K. Wootters {\it  Int.J. Theor Phys } {\bf D 23}, 701, (1984);
   See  also   ``Time in Quantum Gravity:  An Hypothesis'' 
C. Rovelli {\it  Phys. Rev.} {\bf D 43}, 442, (1991)  and references  therein.

\bibitem{Gambini-Pullin} 
``Realistic Clocks, Universal Decoherence and the Black Hole information paradox"
 R. Gambini, R. A. Porto, J. Pullin,
 {\it Phys. Rev. Lett.} {\bf 93}, 240401, (2004),  
[e-Print: hep-th/0406260];
``Fundamental Decoherence  from relational Time in Discrete Quantum Gravity: Galilean Covariance"
R. Gambini, R. A. Porto, J. Pullin, {\it  Phys. Rev.} {\bf D 70}, 124001,(2004), 
[e-Print: gr-qc/0408050].


 \bibitem{Diosi} ``A universal master equation for the gravitational violation of quantum mechanics" 
  L. Diosi,  {\it Phys. Lett. A} {\bfÊ120}, 377, (1987);  ``Models for universal reduction of macroscopic quantum fluctuations",  L. Diosi,  {\it Phys. Lett. A} {\bfÊ40}, 1165,(1989);  
  ``Gravitation and quantum mechanical localization of macro-objects", L. Di\'osi:  {\it Phys.Lett. } {\bf  105A} , 199-202 (1984);
 Gravitation and Quantum Mechanics of macroscopic Objects" F. Karolyhazy,
 {\it Nouvo Cimento}, {\bf XLII}, 1506 (1966).


\bibitem {Otro}
P. M. Pearle in ``Open
Systems and Measurement in Relativistic Quantum Theory",
F. Petruccione and H.P. Breuer eds. (Springer Verlag,1999).
e-Print:quant-ph/9901077;
 ``Combining Stochastic Dynamical State Vector Reduction With
 Spontaneous Localization" P. M. Pearle, 
 {\it Phys. Rev. A} {\bf 39}, 2277,  (1989);
``Reduction of the State Vector by a Nonlinear Schroedinger
Equation." P. M. Pearle {\it Phys. Rev. D} {\bf 13}, 857, (1976).


\bibitem{Bohm} The  classic  reference is {\it Phys. Rev.} {\bf 85}, 166, (1952), although the first proposal was in fact  made in
 by L. de Broglie at the  1927 Solvay Conference, and  reported   in {\it \'Electrons et Photons: Rapports  et Discussions  du Cinquieme Conseil  de Physique} (Gauthier-Villars, Paris. 1928).


\bibitem{Valentini}`` Inflationary Cosmology as a Probe of Primordial Quantum Mechanics"
A. Valentini, {\it Phys. Rev. D} {\bf 82}, 063513, (2010).


\bibitem{ConsitentHistories} Consistent interpretation of quantum mechanics using quantum trajectories.
R. B. Griffiths, 
 {\it Phys. Rev. Lett.} {\bf70},2201, (1993);
``Choice of consistent family, and quantum incompatibility"
R. B. Griffiths, 
  {\it Phys. Rev. A} {\bf 57},1604, (1998); 
``Consistent interpretations of quantum mechanics".
R. Omnes, 
  {\it Rev. Mod. Phys.} {\bf 64}, 339, (1992);
``Quantum classical correspondence using projection operators"
R. Omnes, 
  {\it J. Math. Phys.} {\bf 38}, 697, (1997);
``Classical equations for quantum systems"
M. Gell-Mann, J. B. Hartle, 
 {\it   Phys.Rev. D } {\bf 47}, 3345,(1993). 
e-Print: gr-qc/9210010;
``Comment on `Consistent sets yield contrary inferences in quantum theory'".
R. B. Griffiths,  J. B. Hartle, 
 {\it  Phys. Rev. Lett.} {\bf 81},1981, (1998); 
`` Alternative decohering histories in quantum mechanics"
M. Gell-Mann, J. B. Hartle,  
 Singapore H. E. Phys.1990:1303-1310 (QCD161:H51), (1990).
 
 
  
\bibitem{Laura}`Time Translation of Quantum Properties"
R. Laura \& L. Vanni,
  {\it  Foundations of Physics} {\bf 39},160, (2009)
  
\bibitem{Laura2} Private communication with R. Laura.
  
  
\bibitem{HartleCosmology} 
``Generalizing quantum mechanics for quantum space-time"
J. B. Hartle,Contributed to 23rd Solvay Conference in Physics: The Quantum Structure of Space and Time, Brussel, Belgium, 1-3 Dec 2005. 
Published in *Brussels 2005, The quantum structure of space and time* 21-43 
e-Print: gr-qc/0602013;
 ``Generalizing quantum mechanics for quantum gravity"
J. B. Hartle, 
 {\it Int. J. Theor. Phys.}  {\bf 45}, 1390, (2006); 
``The Quasiclassical realms of this quantum universe"
 J.  B. Hartle. e-Print: e-Print:08;
``The Classical Universes of the No-Boundary Quantum State"
J. B. Hartle, S.W. Hawking, 
  {\it Phys. Rev. D} {\bf 77}, 123537, (2008).
  
  \bibitem{HartleCosmologyNew}
   ``The No-Boundary Measure in the Regime of Eternal Inflation",
 J. B. Hartle , S. W. Hawking, T. Hertog, e-Print: e-Print:1001.0262 [hep-th].

\bibitem{Hartle-Languaje}
``Quantum Physics  and Human Language" J. B.  Hartle,  {\it  J. Phys.  A} {\bf 40}, 3101, (2007).


\bibitem{CHCritics}
``On the consistent histories approach to quantum mechanics"
F.  Dowker \& A. Kent,   {\it J. Statist. Phys.} {\bf 82}, 1575, (1996). 
e-Print: gr-qc/9412067;
 ``Can the decoherent histories description of reality be considered satisfactory?"
A. Bassi \& G. C. Ghirardi, 
{\it Phys. Lett.  A} {\bf 257}, 247, (1999). 
e-Printe. gr-qc/9811050;
 ``About the notion of truth in the decoherent histories approach: A Reply to Griffiths".
A. Bassi \& G. C. Ghirardi,
{\it Phys. Lett. A} {\bf 265}, 153, (2000). 
e-Printe: quant-ph/9912065


\end{thebibliography}
\end{document}